\documentclass[twocolumn,preprintnumbers,amssymb,amsmath,superscriptaddress,letterpaper,nofootinbib,showpacs]{revtex4}[10pt]

\usepackage{graphicx}
\usepackage{dcolumn}
\usepackage{bm}
\usepackage{natbib}
\usepackage{calligra}
\usepackage[T1]{fontenc}
\usepackage{egothic}
\usepackage[T1]{fontenc}
\newfont{\rsfsten}{rsfs10 scaled 1200}
\newfont{\rsfsseven}{rsfs10 scaled 1200}
\newfont{\rsfsfive}{rsfs10 scaled 1200}
\usepackage{epsfig}
\usepackage{units}

\newcommand{\be}{\begin{equation}}
\newcommand{\ee}{\end{equation}}
\newcommand{\bea}{\begin{eqnarray}}
\newcommand{\eea}{\end{eqnarray}}

\def\lsim{\mathrel{\raise.3ex\hbox{$<$\kern-.75em\lower1ex\hbox{$\sim$}}}}
\def\gsim{\mathrel{\raise.3ex\hbox{$>$\kern-.75em\lower1ex\hbox{$\sim$}}}}

\begin{document}

\hspace*{130mm}{\large \tt FERMILAB-PUB-13-547-A}
\vskip 0.2in

\title{Constraining the origin of the rising cosmic ray positron fraction with the boron-to-carbon ratio}

\author{Ilias Cholis}
\email{cholis@fnal.gov}
\affiliation{Fermi National Accelerator Laboratory, Center for Particle Astrophysics, Batavia, Illinois, 60510, USA}
\author{Dan Hooper}
\email{dhooper@fnal.gov}
\affiliation{Fermi National Accelerator Laboratory, Center for Particle Astrophysics, Batavia, Illinois, 60510, USA}
\affiliation{University of Chicago, Department of Astronomy and Astrophysics, Chicago, Illinois, 60637, USA}
\date{\today}

\begin{abstract}

The rapid rise in the cosmic ray positron fraction above 10 GeV, as measured by \textit{PAMELA} and \textit{AMS}, suggests the existence of nearby primary sources of high energy positrons, such as pulsars or annihilating/decaying dark matter. In contrast, the spectrum of secondary positrons produced through the collisions of cosmic rays in the interstellar medium is predicted to fall rapidly with energy, and thus is unable to account for the observed rise.  It has been proposed, however, that secondary positrons could be produced and then accelerated in nearby supernova remnants, potentially explaining the observed rise, without the need of primary positron sources. Yet, if secondary positrons are accelerated in such shocks, other secondary cosmic ray species (such as boron nuclei, and antiprotons) will also be accelerated, leading to rises in the boron-to-carbon and antiproton-to-proton ratios. The measurements of the boron-to-carbon ratio by the \textit{PAMELA} and \textit{AMS} collaborations, however, show no sign of such a rise. With this new data in hand, we revisit the secondary acceleration scenario for the rising positron fraction. Assuming that the same supernova remnants accelerate both light nuclei (protons, helium) and heavier cosmic ray species, we find that no more than $\sim$25\% of the observed rise in the positron fraction can result from this mechanism (at the 95\% confidence level).

\end{abstract}


\maketitle

\section{Introduction}
\label{sec:intro}

Recently, the \textit{AMS} collaboration reported their measurement of the cosmic ray (CR) positron fraction over the range of 0.5 to 350 GeV~\cite{AMS02}, confirming with greater precision the rise at energies above $\sim$10 GeV~\cite{Adriani:2008zr} as previously observed by the \textit{PAMELA}~\cite{Picozza:2006nm} and \textit{Fermi}~\cite{Gehrels:1999ri} collaborations. Proposed explanations for this rise include dark matter (DM) particles annihilating or decaying in the galactic halo~\cite{Bergstrom:2008gr, Cirelli:2008jk, Cholis:2008hb, Barger:2008su, Cirelli:2008pk, Nelson:2008hj, ArkaniHamed:2008qn, Cholis:2008qq, Nomura:2008ru, Yin:2008bs, Harnik:2008uu, Fox:2008kb, Pospelov:2008jd, MarchRussell:2008tu,Chang:2011xn}, nearby pulsars injecting high-energy positrons into the interstellar medium
~\cite{Hooper:2008kg, Yuksel:2008rf, Profumo:2008ms, Malyshev:2009tw, Grasso:2009ma}, and nearby supernova remnants (SNRs) accelerating secondary
positrons produced in the hadronic interactions of CR protons or nuclei \cite{Blasi:2009hv, Mertsch:2009ph, Ahlers:2009ae}.     

Although annihilating dark matter particles have been shown to be able to account for the observed rise in the positron fraction, such scenarios are quite constrained at this time. In particular, the only dark matter models that can accommodate both the positron fraction and measurements of the electron-plus-positron spectrum \cite{Ackermann:2010ij, Abdo:2009zk, Collaboration:2008aaa, Aharonian:2009ah}, feature dark matter particles with masses of $\sim$1-3 TeV that annihilate to intermediate states which subsequently decay to muons or charged pions \cite{Cholis:2013psa, Cirelli:2008pk}. 
Such a class of models is that of eXciting Dark Matter \cite{Finkbeiner:2007kk}, which can also accommodate the high annihilation rate needed to generate the observed positron fraction~\cite{Cholis:2013psa, Cholis:2008qq, Cholis:2008wq, Chen:2008qs, Nardi:2008ix}
through Sommerfeld enhancements~\cite{ArkaniHamed:2008qn} (see also Refs.~\cite{Fox:2008kb,Lattanzi:2008qa,Hisano:2004ds})\footnote{The presence of near-by DM clumps can not solely explain the necessary high annihilation rate, but can reduce the needed annihilation cross-section by a factor of $\sim 2$ compared to case where no DM clumps are invoked \cite{Kamionkowski:2010mi}.}. For the case of decaying dark matter, models of dynamical dark matter have also been recently proposed in connection with the leptonic data \cite{Dienes:2013lxa}. We also note that even if dark matter does not account for the rising positron fraction, such measurements can be used to derive stringent constraints on dark matter models with mass up to 350 GeV, which annihilate or decay to leptonic final states~\cite{Bergstrom:2013jra, Ibarra:2013zia}. 
 
Pulsars (rapidly spinning neutron stars which steadily convert their rotational kinetic energy into radio emission, gamma-rays, and high-energy electron-positron pairs) could also account for the observed rise in the positron fraction~\cite{Hooper:2008kg, Yuksel:2008rf, Profumo:2008ms, Malyshev:2009tw}.  
In addition to the combined contribution from all pulsars throughout the Milky Way, the young and nearby Geminga and  B0656+14 pulsars could each contribute significantly to the cosmic ray positron spectrum~\cite{Linden:2013mqa, Cholis:2013psa} (see also Ref.~\cite{Gaggero:2013rya}). 

Both dark matter and pulsar origins for the rising positron fraction represent scenarios in which the positrons are cosmic ray primaries. In contrast, it has also been proposed that the excess positrons could be cosmic ray secondaries, produced in proton-proton collisions inside of SNRs and then accelerated before escaping into the interstellar medium (ISM)~\cite{Blasi:2009hv, Mertsch:2009ph}. It is this case that we consider in this study.  In particular, within this scenario, the same stochastic acceleration processes which accelerate CR positrons in the supernova shocks will also accelerate other species of CR secondaries, such as antiprotons and boron nuclei. Thus, as was shown in 
Refs.~\cite{Blasi:2009bd, Mertsch:2009ph}, a rise in the antiproton-to-proton and boron-to-carbon ratios are also expected to occur at high energies, $\gsim 100$ GeV (see though \cite{Kachelriess:2011qv, Kachelriess:2012ag}).
Recently, the \textit{PAMELA} \cite{PAMELABtoC} and \textit{AMS} \cite{AMSsite} collaborations presented their first measurements of the boron-to-carbon ratio, revealing no evidence for any rise up to the highest measured energies, $\sim$400 GeV. In this paper, we make use of this measurement to place constraints on models in which the observed rise in the CR positron fraction is the result of the acceleration of positron secondaries.

The remainder of this article is structured as follows. In Sec.~\ref{sec:set_up}, we describe our calculations of the boron-to-carbon ratio, the antiproton-to-proton ratio, and the positron fraction in some detail. We then present our results in Sec.~\ref{sec:results}.  We find that secondary acceleration models capable of explaining the observed positron fraction are also incompatible with the boron-to-carbon ratio, as measured by \textit{AMS} and \textit{PAMELA}. We summarize our results and conclusions in Sec.~\ref{sec:Conclusions}. 
                   
\section{Calculation Setup and Assumptions}
\label{sec:set_up}

Diffusive shock acceleration in galactic SNRs can be responsible for the spectrum of CRs up to $\sim$ PeV energies (at much higher energies, extragalactic sources are presumably responsible). 
Ambient electrons, protons, and nuclei are accelerated by the shock front, generating a spectrum that is expected to take a power-law form, $dN/dE \propto E^{-\gamma + 2}$, where the index $\gamma$ depends on the conditions of the shock. For a supersonic 
shock the compression ratio, $r= v^{-}/v^{+}$, is taken to be $r=4$, where $v^{+}$ is the plasma down-stream velocity (inside the shock) and $v^{-}$ the plasma up-stream velocity (outside the shock) (both defined in the frame of the shock front). The index $\gamma$ is related to $r$ by $\gamma =3r/(r-1)$. For $r=4$, this yields a $E^{-2}$ injection 
spectrum for the primary CR component. 

While being accelerated inside of the supernova shock, these particles may also interact with the dense gas and spallate or decay to produce lighter 
species \cite{Blasi:2009hv, Mertsch:2009ph}. The relevant source term for these lighter species is given by: 
\begin{equation}
Q_{i}(E_{kin}) = \Sigma_{j}N_{j}(E_{kin}) \left[ \sigma^{sp}_{j \rightarrow i} \, \beta \, c \, n_{gas} + \frac{1}{E_{kin}\,\tau^{dec}_{j \rightarrow i}} \right],
\label{eq:sourceTerm}
\end{equation}
where $E_{kin}$ is the kinetic energy per nucleon (in GeV), $N_{j}$ gives the spectrum of the parent nucleus species $j$, $\sigma^{sp}_{j \rightarrow i}$ 
is the partial cross section from species $j$ to species $i$, $\tau^{dec}_{j \rightarrow i}$ is the timescale for the decay of species $j$ to $i$, and $n_{gas}$ is the density of gas where the spallation occurs.

The same processes also provide a corresponding loss term: 
\begin{equation}
\Gamma_{i}(E_{kin}) =  \sigma^{sp}_{i} \, \beta \, c \, n_{gas} + \frac{1}{E_{kin}\,\tau^{dec}_{i}}, 
\label{eq:CRLossRate}
\end{equation}
where $\sigma^{sp}_{i}$ and $\tau^{dec}_{i}$ are the total spallation cross section and total decay lifetime of nuclei species $i$, respectively.

Combining Eqs.~\ref{eq:sourceTerm} and~\ref{eq:CRLossRate} with the effects of advection, diffusion, and adiabatic energy losses, one gets the transport equation for species $i$:
\begin{equation}
v \frac{\partial f_{i}}{\partial x} = D_{i}\frac{\partial^{2} f_{i}}{\partial x^{2}} + \frac{1}{3}\frac{d v}{d x}p\frac{\partial f_{i}}{\partial p} - \Gamma_{i}f_{i} + q_{i},
\label{eq:TranspEq}
\end{equation}
where $f_{i}$ is the phase space density of CR species $i$ and $q_{i}$ is the relevant source term.  
CRs are typically accelerated in the shock over a timescale on the order of $\tau^{SN}$ $\sim 10^{4}$ yr. If enough nuclei of species $i$ are produced via spallation or decay, and are accelerated in the SNR before undergoing further spallation or decay (1/$\Gamma_{i}$ $\gg$ $\tau^{acc}$),
this can have a significant impact on the CR spectrum. The additional component resulting from this process is referred to as the secondary CRs accelerated inside of the SNRs. The authors of Ref.~\cite{Mertsch:2009ph} solved Eq.~\ref{eq:TranspEq} analytically and calculated the phase space densities for particles, $i$, both up-stream and down-stream from the shock front, including both primary and secondary accelerated CRs. Here, we will use the same formalism, and present portions of their calculation where necessary (see Ref.~\cite{Mertsch:2009ph} for more details).

In solving Eq.~\ref{eq:TranspEq}, we apply the boundary conditions that the phase space density for species $i$ far up-stream (far away from the supernova shock) is equal to the 
ambient density $Y_{i}$, and its gradient in momentum is zero:  
\begin{eqnarray}
&&\lim_{x\rightarrow - \infty} \,\, f_{i}(x,p) = Y_{i}\delta (p - p_{0}), \\ 
&&\lim_{x\rightarrow - \infty} \frac{\partial f_{i}(x, p)}{\partial p} = 0. \nonumber
\label{eq:farUpstream}
\end{eqnarray}
Following Ref.~\cite{Mertsch:2009ph}, the the phase space density  down-stream, $f^{+}_{i}$, is given by:
\begin{equation}
f_{i}^{+}(x,p) = f_{i}(0,p) + \frac{q^{+}_{i}(0,p)-\Gamma_{i}^{+}(p)f_{i}(0,p)}{v^{+}}x, 
\label{eq:Downstream}
\end{equation}
where $x$ is the distance from the shock front and $q_i^{\pm}$ is the total source term for species $i$, given by: 
\begin{equation}
q_{i}^{\pm}(x,p) = \Sigma_{j>i} \, f_{j} \, \Gamma^{\pm}_{j \rightarrow i}.
\label{eq:UpDownSource}
\end{equation}

The only difference between $\Gamma^+$ and $\Gamma^-$ comes from different down-stream and up-stream gas densities. 
Ignoring the decay lifetimes of CRs inside and around the supernova shock, we get that $q^{+}_{i}/q^{-}_{i}$ = $\Gamma^{+}_{i}/\Gamma^{-}_{i}$ 
= $n^{+}_{gas}/n^{-}_{gas} = r$. Following Ref.~\cite{Mertsch:2009ph}, we also assume that $D^{+}_{i} = D^{-}_{i}$. 

Integrating the transport equation over infinitesimal distance one gets \cite{Mertsch:2009ph}:
\begin{eqnarray}
p\frac{\partial f_{i}(x, p)}{\partial p} &=&  - \gamma f_{i}(0, p) - \gamma(1+r^{2}) \frac{\Gamma^{-}_{i}(p)D^{-}_{i}(p)}{(v^{-})^{2}}f_{i}(0, p) \nonumber \\
&+& \gamma \large[ (1+r^{2}) \frac{q^{-}_{i}(0, p)D^{-}_{i}(p)}{(v^{-})^{2}} + Y_{i} \delta (p -p_{0})\large].
\label{eq:transportEq}
\end{eqnarray}
%
The solution to which yields:
\begin{eqnarray}
 f_{i}(0, p)&=& \int^{p}_{0} \frac{d p'}{p'} \bigg(\frac{p'}{p}\bigg)^{\gamma} e^{- \gamma (1 + r^{2})(D^{-}_{i}(p) - D^{-}_{i}(p'))\Gamma^{-}_{i}(p)/(v^{-})^{2}} \nonumber \\
&\times& \gamma \large[ (1+r^{2}) \frac{q^{-}_{i}(0, p')D^{-}_{i}(p')}{(v^{-})^{2}} + Y_{i} \delta (p' -p_{0})\large].
\label{eq:transportEq2}
\end{eqnarray}
Following Refs.~\cite{Blasi:2009hv, Mertsch:2009ph}, we assume Bohm diffusion for CRs around the shock front:
\begin{equation}
D_{i}^{\pm}(E) = \frac{K_{B}\, r_{L}(E)\,c}{3} = 3.3 \times 10^{22} \,K_{B} \, B^{-1} \, E \, Z_{i}^{-1}   {\rm cm}^{2} \,{\rm s}^{-1},
\label{eq:BohmDiff}
\end{equation}
where $r_{L}$ is the Larmor radius around the shock front, $B$ is the magnetic field in $\mu G$, $Z$ the atomic number of the CR nucleus $i$, and $E$ is the energy in GeV. $K_{B}$ is a ``fudge factor'' \cite{Mertsch:2009ph} which scales approximately as $K_{B} \simeq (B/\delta B )^{2}$ \cite{Blasi:2009hv}, allowing for faster diffusion of CRs around the shock front. Values of $K_{B} >> 1$ have been suggested~\cite{Blasi:2009hv, Mertsch:2009ph} under conditions where magnetic field amplification is inefficient. 

\begin{figure}
\includegraphics[width=3.40in,angle=0]{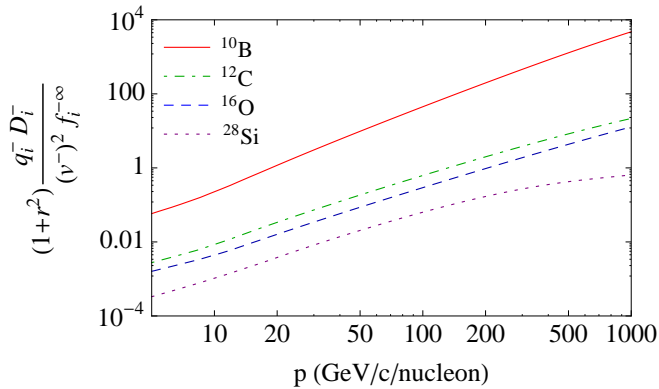}
\caption{The ratio of the secondary cosmic ray acceleration term to the primary cosmic ray acceleration term in Eqs.~\ref{eq:transportEq} and~\ref{eq:transportEq2}, 
as a function of momentum per nucleon. The impact of including the acceleration of secondary cosmic rays produced inside and around the supernova shock 
front is most important at high energies and for lighter species. For $^{10}$B the ratio is significantly higher since $f^{-\infty}_{^{10}\textrm{B}}$ is suppressed. As in Ref.~\cite{Mertsch:2009ph}, we have adopted optimistic values for $K_{B}$, $B$, $n^{-}_{gas}$, $v^{-}$ and $r$ (see text for more details).}
\label{fig:sec_to_prim}
\end{figure}

The importance of including the acceleration of secondary CRs produced inside and around the supernova shock front varies with energy and CR species. 
In Fig.~\ref{fig:sec_to_prim} we show the ratio of the secondary CR acceleration term of Eqs.~\ref{eq:transportEq} and~\ref{eq:transportEq2}, 
$(1+r^{2}) q^{-}_{i}(0, p)D^{-}_{i}(p)/(v^{-})^{2}$, to the primary CR acceleration term, $Y_{i} \delta (p -p_{0}) = f^{-\infty}_{i}$, as a function of momentum per nucleon. This ratio increases with energy and is greater for the lighter species. This demonstrates that the acceleration of CR secondaries is most important in the case of light nuclear species, and at high energies. As in Ref.~\cite{Mertsch:2009ph}, we have adopted the following parameter values: $K_{B} = 40$, $B = 1$ $\mu$G, $n^{-}_{gas} = 2$ cm$^{-3}$, $v^{-} = 0.5 \times 10^{8}$ cm$^{3}$\,s$^{-1}$, and $r=4$, which have been suggested from the observed titanium-to-iron ratio, and are also similar to those proposed from the positron fraction ($K_{B}$ = 20 \cite{Blasi:2009hv}). 


%

We calculate the far up-stream phase space densities from the measured CR densities for Fe, Si, Mg Ne, 
O, N, C, B, He, and p \cite{Engelmann:1990zz, AMSsite}, taking into account the relative isotopic abundances. 
For the calculation of the boron-to-carbon ratio, we start from $^{18}$O and go down to $^{10}$Be.\footnote{$^{10}$Be decays to $^{10}$B with a lifetime of 1.36 Myr.} We employ the relevant total and partial cross sections (see Refs.~\cite{Silberberg:1990nj, Silberberg:1973jxa}). We then use the same formulation to calculate the antiproton-to-proton ratio, and the positron fraction, including helium and proton CRs. We start from the heaviest isotope and solve Eqs.~\ref{eq:UpDownSource} and~\ref{eq:transportEq2} to obtain 
the injected spectrum of CRs after integrating over the volume of the SNR:
\begin{equation}
N_{i}(E) = 16 \pi^2 \int^{v^{+} \tau^{SN}}_{0} dx \, p^{2} f^{+}_{i}(x, p)  \, (v^{+} \tau^{SN} - x)^2.
\label{eq:SNR_InjSpect}
\end{equation}
We take $\tau^{SN} = 2\times10^{4}$ yr and $v^{+} = 1.25\times10^{7}$ cm s$^{-1}$.

Once CRs are injected into the ISM, they propagate in the galactic medium. Depending on the CR species and their energy scale, 
there are various possibly relevant time-scales. The CR diffusion, the CR advection, the diffusive re-acceleration time-scales,
the decay time-scale and the total energy losses time-scale. In addition, as we stated earlier, CR secondaries are produced in
the interstellar medium. Depending on the aimed level of accuracy and which are the important time-scales, one can solve the
 propagation equation for CRs analytically, including only diffusion and advection (see \cite{LorimerV1}), use a leaky box approximation (as we do), 
 or solve numerically including all effects \cite{Galprop1,website,DRAGONweb}. For CR protons, anti-protons, Boron and Carbon and for the energies at hand, advection, re-acceleration and energy losses in the interstellar medium are subdominant (for CR electrons and positrons energy losses have to be included). 
These CRs diffuse within a zone of scale height $L \sim$ 1\,-\,8 kpc~\cite{Simet:2009ne, Trotta:2010mx}, beyond which they are free to escape. The escape timescale for a CR nucleus $i$ is $\tau^{esc}_{i}(E) \simeq \tau^{esc}_{1} \times (E/Z)^{-\delta}$, where $E$ is in GeV, $Z$ is the atomic number, and $\delta$ is the diffusion index. The normalization, $\tau^{esc}_{1}$, and the index, $\delta$, can be extracted by fitting
the boron-to-carbon ratio at energies below $\sim 30$ GeV, where the effects of the acceleration of secondaries inside SNRs are subdominant. 

The density of CR nuclei at Earth (neglecting solar modulation) is given by:
\begin{equation}
\textrm{\rsfsten{N}}_{i}(E) = \frac{\Sigma_{j>i} (\Gamma^{sp}_{j \rightarrow i} 
+ 1/(E_{kin}\tau^{dec}_{j \rightarrow i}))\textrm{\rsfsten{N}}_{j}(E) + \textrm{\emph{$R_{SN}$}}N_{i}(E)}{\Gamma_{i}(E) + 1/\tau^{esc}_{i}(E)}.
\label{eq:ISM_Spect}
\end{equation}
$\textrm{\emph{$R_{SN}$}}$, is the galactic supernovae rate per volume (3 per century in the galactic disk). 
%
%
For secondary electrons and positrons produced in p-p and He-p collisions and then further accelerated inside of the SNR, there are no decay or spallation process to take into account ($\Gamma^{\pm}_{e} = 0$), and thus Eq.~\ref{eq:transportEq2} simplifies to: 
\begin{equation}
 f^{acc}_{e^{\pm}}(0, p)=\int^{p}_{0} \frac{d p'}{p'} \bigg(\frac{p'}{p}\bigg)^{\gamma} \gamma (\frac{1}{\xi}+r^{2}) \frac{q^{-}_{e^{\pm}}(0, p)D^{-}_{e}(p')}{(v^{-})^{2}}.
\label{eq:transportEq2pos}
\end{equation}
We take $\xi=$0.05, as about 5\% of the energy of the primary CR proton goes into each $e^{\pm}$ in an inelastic p-p collision.\footnote{For CRs other than protons and antiprotons from p-p or p-He collisions, $\xi$ is taken to be 1 (the daughter 
particle produced from spallation has, on average, about the same momentum per nucleon as the CR primary).}

Once released into the ISM, CR electrons and positrons undergo diffusion and energy loss processes. In this work, we focus on energies above 5 GeV where solar modulation effects are small. 
Above a few GeV, the $e^{\pm}$ energy losses are dominated by a combination of synchrotron and inverse Compton scattering. 
We model these energy losses as $dE_{e}/dt = b(E)$$=b_{0} (E_{e}/1\,{\rm GeV})^{2}$, with $b_{0} = -1.7 \times 10^{-16}$ GeV s$^{-1}$. 
The value for the $b_{0}$ coefficient comes from estimates on the local magnetic and radiation fields.\footnote{We assume a local magnetic field value of $B=5 \mu$G, corresponding to an energy density of $U_{mag}=$0.62 eV/cm$^3$. For the local radiation energy density, we take $U_{rad}=$0.82 eV/cm$^3$ for the galactic radiation field and $U_{CMB}=$0.26 eV/cm$^3$ for the cosmic microwave background. The energy loss rate is $\frac{dE}{dt} \simeq -\frac{4}{3}\sigma_{T}c \gamma^{2}$$(U_{mag}+U_{rad}+U_{CMB})$ (where $\gamma$ here is the Lorentz boost).}
The escape timescale for electrons, $\tau^{esc}_{e}(E)$, is the same as that for protons, as at 
high energies ($E \gg m$) they have the same rigidity. Their steady state density is given by: 
%
\begin{equation}
\textrm{\rsfsten{N}}^{\; \; acc}_{e^{\pm}}(E) =   \textrm{\emph{$R_{SN}$}} \frac{1}{b(E) + 1/\tau^{esc}_{e}(E)} \int^{E_{max}}_{E} dE' N^{acc}_{e^{\pm}}(E'),
\label{eq:ISM_Spect_SecAccPos}
\end{equation}
where $N^{acc}_{e^{\pm}}(E')$ is calculated by replacing $f^{+}_{i}(x, p)$ with  $ f^{acc}_{e^{\pm}}(x>0, p)$ in Eqs.~\ref{eq:Downstream} and~\ref{eq:SNR_InjSpect}.
We include both CR protons and CR helium nuclei in the source term of Eq.~\ref{eq:transportEq2pos}. 
Any additional correction factor in Eq.~\ref{eq:ISM_Spect_SecAccPos} due to the impact of heavier CR species is expected to be at the level of $\sim$$10\%$, and is highly dependent on the initial chemical composition of the surrounding medium. 

Primary CR electrons (due to the $Y_{e^{-}} \delta (p - p_{0})$ term not included in Eq.~\ref{eq:transportEq2pos})\footnote{In this paper we ignore the presence of positrons in 
the ambient interstellar medium. That is a simplification since we have observed positrons at many different energies. Yet their ratio to electrons is not known at energies much lower than 0.5 GeV.}
are given by:
\begin{eqnarray}
\textrm{\rsfsten{N}}^{\; \; prim}_{e^{-}}(E) &=& K_{e^{-}} \textrm{\emph{$R_{SN}$}} \frac{1}{b(E) + 1/\tau^{esc}_{e}(E)} \nonumber \\
&\times& \int^{E_{max}}_{E} dE' N^{prim}_{e^{-}}(E'),
\label{eq:ISM_Spect_PrimElec}
\end{eqnarray}
where $E_{max}$ is the maximum energy to which $e^{\pm}$ particles are accelerated. We set the value of $K_{e^{-}}$ to match local CR measurements.

Finally, CR $e^{\pm}$ are also produced in p-p and He-p collisions in the ISM. We ignore diffusive re-acceleration in the ISM, as its impact is expected to be subdominant above a few GeV.
The secondary CR $e^{\pm}$ flux is given by:
\begin{eqnarray}
\label{eq:ISM_Spect_SecPos}
\textrm{\rsfsten{N}}^{\; \; sec}_{e^{\pm}}(E) &=& n_{ISM} c \frac{1}{b(E) + 1/\tau^{esc}_{e}(E)}  \int^{E_{max}}_{E} dE''  \\
& & \int^{\infty}_{5 E''} dE' \Sigma_{i} \textrm{\rsfsten{N}}^{\; \; prim}_{i}(E') \frac{d\sigma_{i \rightarrow e^{\pm}}(E', E'')}{dE'}, \nonumber
\end{eqnarray}
where the sum is carried out over protons and helium nuclei. $E''$ is the energy of the secondary $e^{\pm}$ at production and $E'$ the energy of the parent CRs in the ISM. 
The factor of 5 in the lower limit of integration comes from the fact that a charged pion produced in a p-p collision carries approximately 1/5 of the energy of the parent CR proton.\footnote{Our results do not depend significantly on the precise value of this lower limit of integration.}
 It is the decays of these charged pions that produce the secondary electrons and positrons.

When comparing our results to observations, we include the effects of solar modulation, using the force field approximation 
\cite{Gleeson:1968zza}. More recent models include charge-sign dependent solar modulation 
\cite{2011ApJ...735...83S, 2012Ap&SS.339..223S, Maccione:2012cu} and can impact the positron fraction and the antiproton-proton 
ratios by changing differently the fluxes of electrons(antiprotons) from positrons(protons) of the same energy before entering the 
Heliosphere. Even in those cases though, the effects of solar modulation on the CR ratios is always negligible above 10 GeV (GeV/n) \cite{Maccione:2012cu}.


\begin{figure*}
\begin{centering}
\hspace{-0.5cm}
\includegraphics[width=3.40in,angle=0]{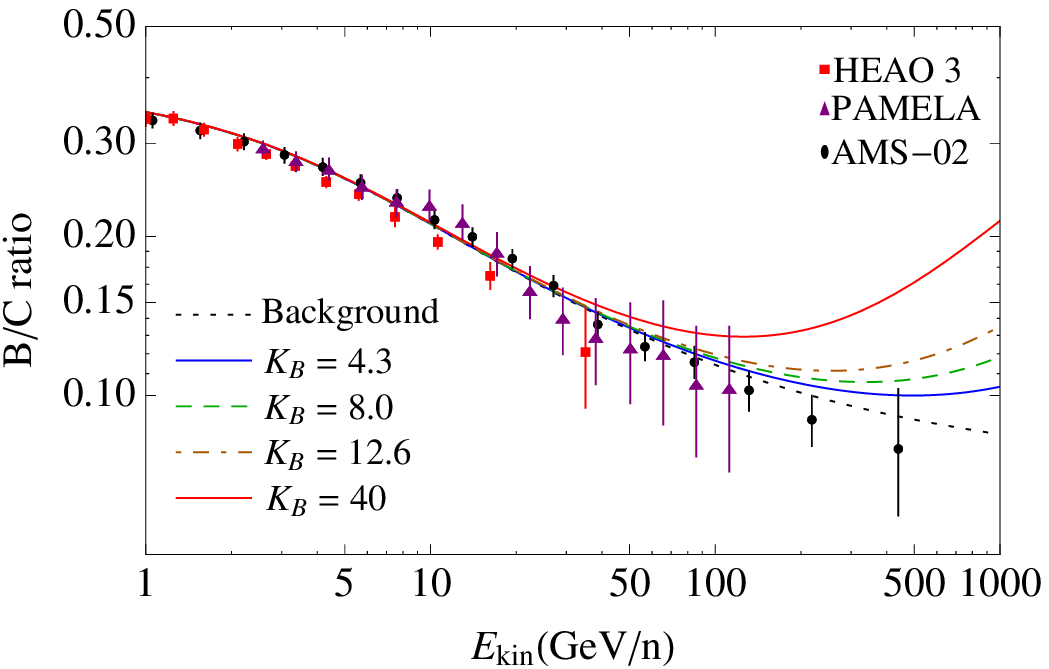}
\includegraphics[width=3.40in,angle=0]{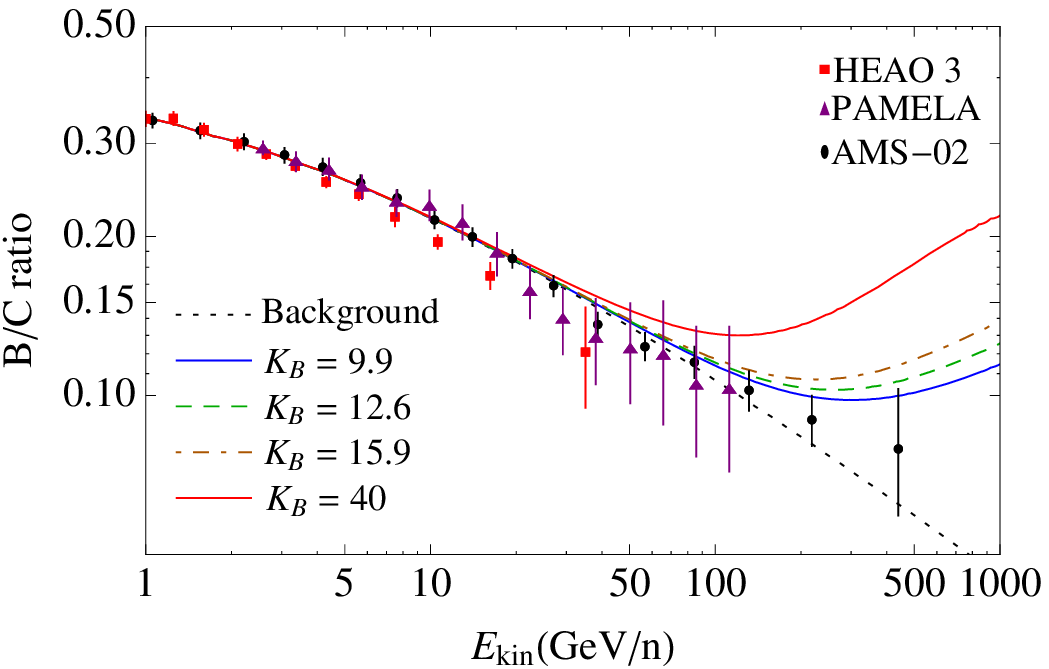} \\
\hspace{-0.5cm}
\includegraphics[width=3.40in,angle=0]{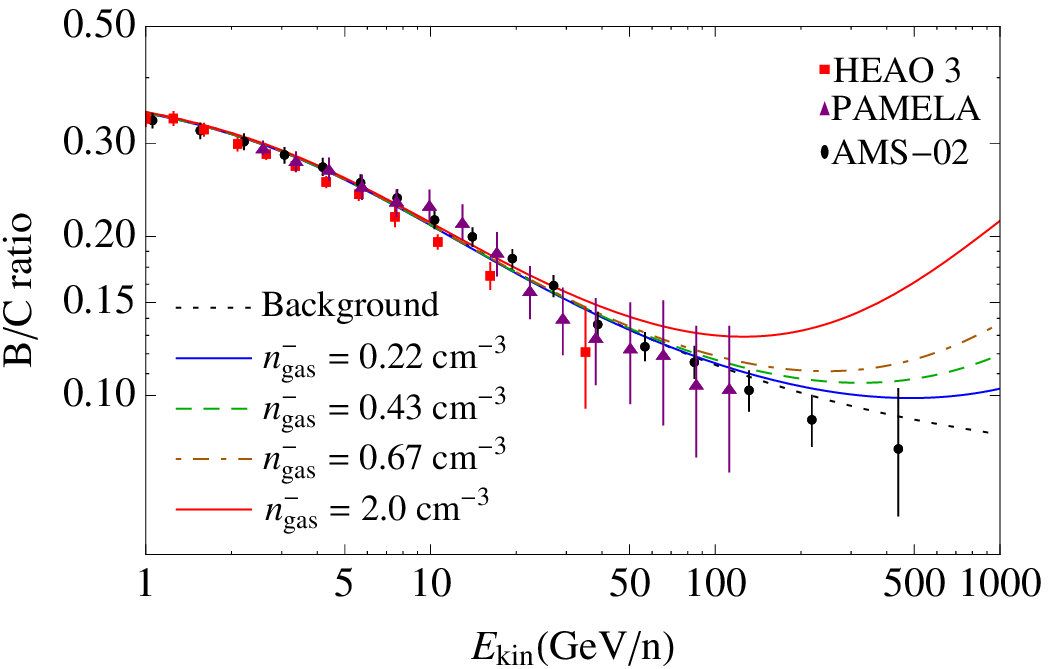}
\includegraphics[width=3.40in,angle=0]{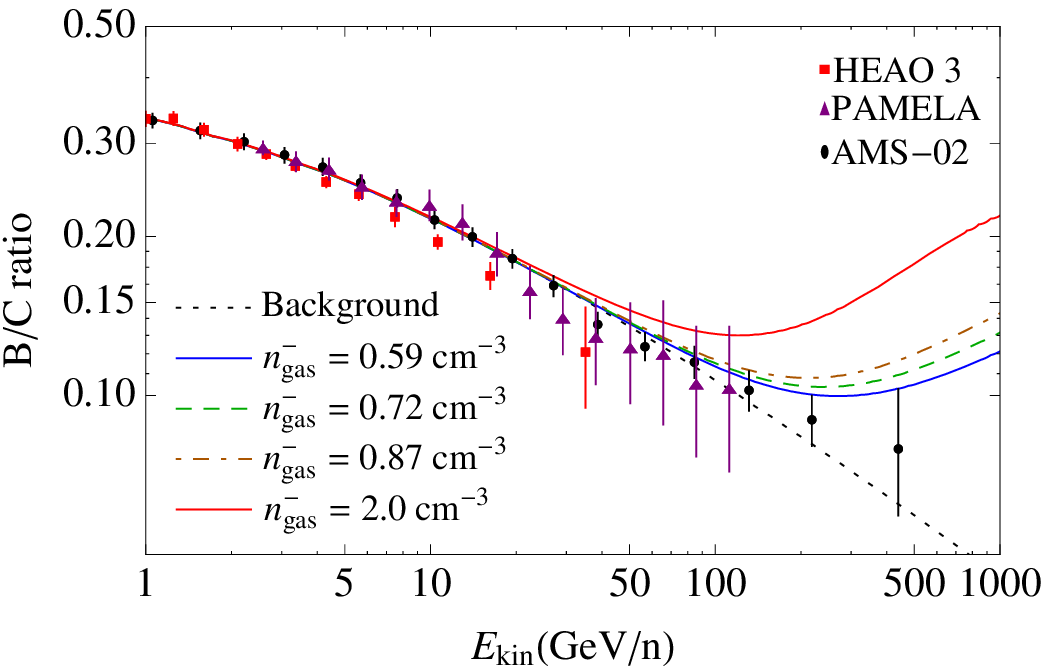}
\end{centering}
\caption{The cosmic ray boron-to-carbon ratio predicted for various parameter choices. In each frame, the black dotted curves represent the prediction without any contribution from the acceleration of secondary CRs in SNR shocks.  In the the left frames, this is calculated according to Eq.~\ref{eq:transportEq2} 
with $q^{-}_{i} = 0$, whereas in the right frames we have used GALPROP (see text for details). The other curves include contributions from accelerated secondaries. In the upper frames, we consider different values of $K_B$, and set $n^{-}_{gas}$ = 2 cm$^{-3}$. In the lower frames, we set $K_B=40$ and vary the value of $n^{-}_{gas}$. In all frames, we set $B$ = 1 $\mu$G, $v^{-}$ = 0.5$\times 10^{8}$ cm s$^{-1}$ and $r$ = 4. In each frame, the solid blue, dashed green, and dashed-dotted brown curves represent parameter choices that are incompatible with the measured boron-to-carbon spectrum at the 95\%, 99\%, and 99.9\% confidence levels, respectively (using the combination of data from \textit{AMS} and \textit{PAMELA}; HEAO 3 data is shown only for comparison). We also include in each frame the prediction for an even more extreme parameter value ($K_B=40$, $n^{-}_{gas}$ =2.0 cm$^{-3}$) for comparison with Ref.\cite{Mertsch:2009ph}.}
\label{fig:B_to_C}
\end{figure*}

\begin{figure*}
\begin{centering}
\hspace{-0.5cm}
\includegraphics[width=3.40in,angle=0]{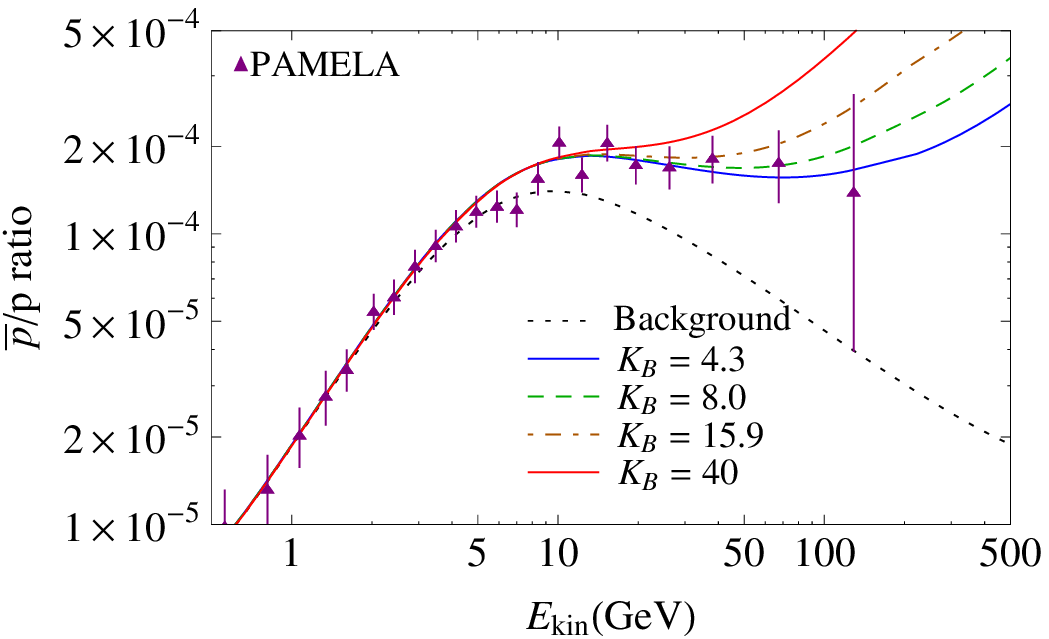}
\includegraphics[width=3.40in,angle=0]{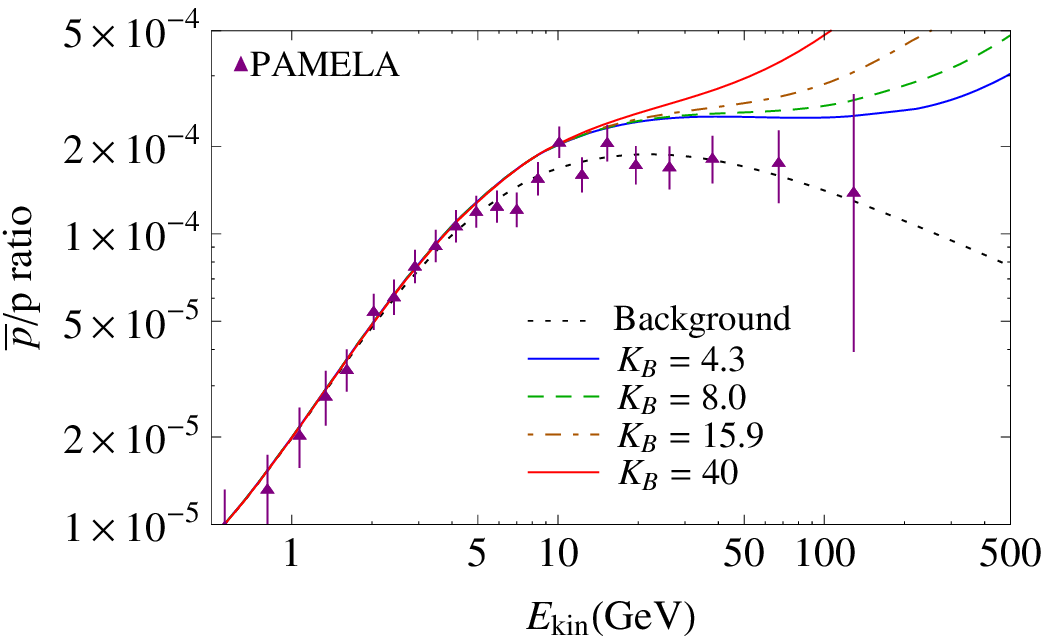} \\
\hspace{-0.5cm}
\includegraphics[width=3.40in,angle=0]{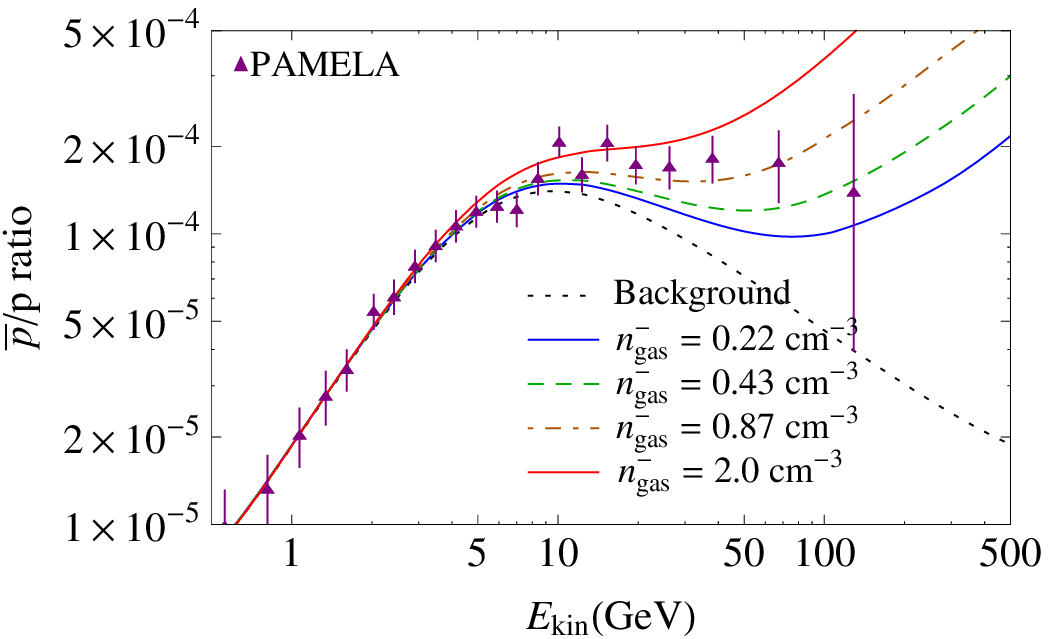}
\includegraphics[width=3.40in,angle=0]{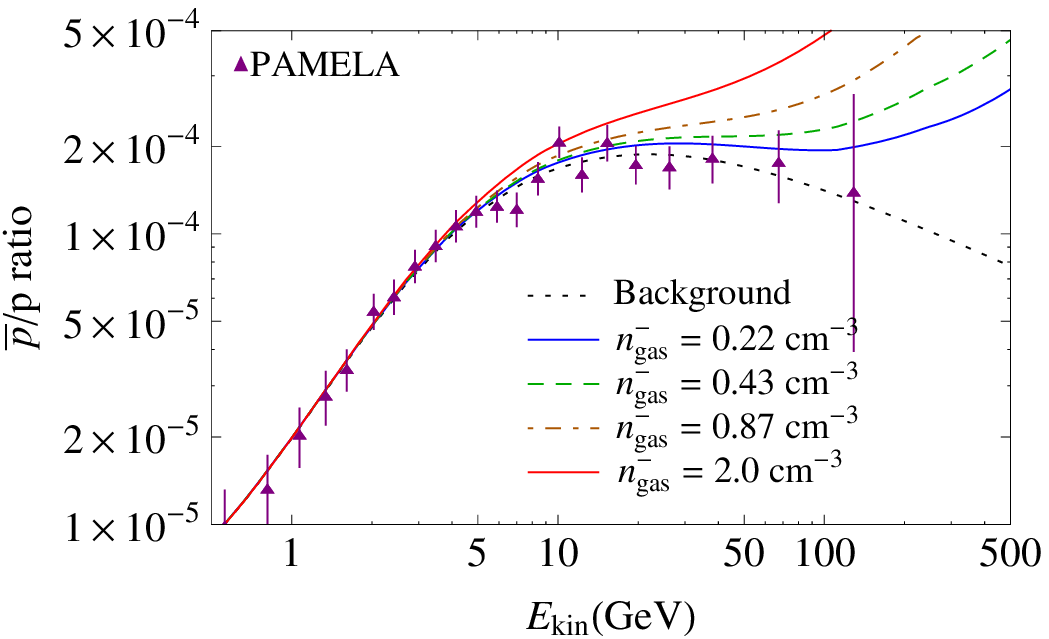}
\end{centering}
\caption{The cosmic ray antiproton-to-proton ratio predicted for some of the same parameter choices used in Fig.~\ref{fig:B_to_C}. See text for details.}
\label{fig:pbar_to_p}
\end{figure*}

\section{Results}
\label{sec:results}

In Fig.~\ref{fig:B_to_C}, we show the CR boron-to-carbon ratio as a function of energy, as predicted for a range of parameter values. In each frame, the dotted black curve denotes the prediction, assuming that secondary cosmic rays are produced only in the ISM and are not subsequently accelerated. In the left frames, this was calculated using Eq.~\ref{eq:transportEq2} with $q^{-}_{i} = 0$, and adopting parameter values of $\delta = 0.6$ and $\tau^{esc}_{1} = 65$ Myr. 
Based on older boron-to-carbon measurements from the High Energy Astrophysics Observatory (HEAO) \cite{Engelmann:1990zz},  the Cosmic-Ray Energetics and Mass experiment (CREAM) \cite{Ahn:2008my}, the Cosmic Ray Nuclei experiment  (CRN)  \cite{1990ApJ...349..625S}, and the Advanced Composition Explorer (ACE) \cite{deNolfo:2006qj}, values of $\tau^{esc}_{1}\sim$ 20 - 80 Myr had been favored~\cite{Blasi:2009hv, Mertsch:2009ph,Trotta:2010mx}. 
In fact, our values for $\tau^{esc}_{1}$ and $\delta$ yield a good fit to the data\footnote{These parameters yield a fit of $\chi^{2}_{tot} \simeq 14$, over 27 degrees of freedom. 
We use all the boron-to-carbon data points from \textit{PAMELA} \cite{PAMELABtoC} and \textit{AMS} \cite{AMSsite} with $E_{k} > 2$ GeV/n, and allow for a modulation potential in the range of 0.5-1.5 GV,
following the force field approximation \cite{Gleeson:1968zza}. Effects of diffusive acceleration or advective winds in the interstellar medium impact energies up to $\sim$ 10 GeV/n but are ignored since the acceleration of secondaries is important only above 50 GeV/n.} up to the highest measured energies, without including any additional contribution from the acceleration of secondary CRs inside of SNRs. In the right frames, we instead use the publicly available code GALPROP v54 (see Refs.~\cite{Galprop1,website}, and references therein) to calculate the boron-to-carbon ratio (again, without any contribution from accelerated secondaries).\footnote{GALPROP includes up-to-date information pertaining to the local interstellar radiation field and the distribution of gas in the Galaxy. It also makes different assumptions regarding the inelastic cross sections (see the discussion on antiprotons). 
 Codes such as GALPROP and DRAGON~\cite{DRAGONweb} assume a simple diffusion zone with free escape boundary conditions. Cosmic rays diffuse within the diffusion zone with a diffusion coefficient $D(R)$=$D_{0} (\frac{R}{3GV})^{\delta}$ ($R$ is the rigidity of the particle) and escape upon reaching any boundary of the zone. We take this zone to be a cylinder, extending a distance $L=4$kpc above and below the galactic plane, and radially 20 kpc from the Galactic Center. We do not include any advective winds, but allow for diffusive re-acceleration with an Alfv$\acute{\textrm{e}}$n speed of 10 km s$^{-1}$. For the parameter values, $\delta =0.43$ and 
 $D_{0}=2.95 \times 10^{28}$ cm$^{2}$s$^{-1}$, we find an excellent fit of $\chi^{2} \simeq 9.9$ over 27 degrees of freedom.} For the purposes of this study, the dotted black curves in Fig.~\ref{fig:B_to_C}, which denote the prediction for the case in which secondary particles are not further accelerated in the shocks of SNRs, represent our ``background'', with respect to which we will later calculate our $\Delta \chi^{2}$ in deriving upper limits on the acceleration of secondary CRs. 


To derive limits on the stochastic acceleration of CR secondaries in SNR shocks, we use the recently released boron-to-carbon ratio data from \textit{PAMELA} \cite{PAMELABtoC} and \textit{AMS} \cite{AMSsite}. In each case, the boron-to-carbon ratio is fitted to match the measurements below 30 GeV, where any contribution from accelerated secondaries is insignificant. In both the left and right frames, the dotted black curves are in good agreement with the data at all energies from \textit{PAMELA} and \textit{AMS}, yielding fits with a $\chi^2$ per degree-of-freedom of 0.50 and 0.35, respectively\footnote{Including measurements from CREAM \cite{Ahn:2008my}, CRN  \cite{1990ApJ...349..625S} and TRACER \cite{Obermeier:2011wm}, does not affect our results.}.  

The most important parameters for our calculation are the magnetic field $B$ (which we take to be fixed at
$1 \mu$G), the shock compression ratio $r$ (which we fix to $r=4$), the up-stream velocity $v^{-}$ (which we fix to $v^{-}$=0.5$\times 10^{8}$ cm s$^{-1}$), 
the up-stream gas density $n^{-}_{gas}$ (which we allow to vary), and the factor $K_{B}$ which is related to the efficiency of diffusion around the shock (which we also allow to vary). For the purposes of our calculations, $K_{B}$ and $B$ are degenerate quantities (see Eq.~\ref{eq:BohmDiff}),
thus we choose to vary only $K_{B}$. Also $K_{B}$, $B$, $n^{-}_{gas}$, $v^{-}$ and $r$ are connected since they all appear in the secondary
CR acceleration term of Eqs.~\ref{eq:transportEq} and~\ref{eq:transportEq2}, $(1+r^{2})q^{-}(0,p)_{i}D^{-}_{i}(p)/(v^{-})^{2}$ (see also Eqs.~\ref{eq:CRLossRate},~\ref{eq:Downstream} and~\ref{eq:UpDownSource}). For this reason, we also choose to vary the value of $n^{-}_{gas}$.

In Fig.~\ref{fig:B_to_C}, we show the predicted boron-to-carbon ratio, including the contribution from secondaries produced and accelerated in SNRs, for a range of parameter values. In the upper frames, we set $n^{-}_{gas} = 2$ cm$^{-3}$ and vary $K_{B}$, while in the lower frames we set $K_{B}=40$ and consider different values of $n^{-}_{gas}$. In each frame, the solid blue, dashed green, and dashed-dotted brown curves denote the parameter values which are incompatible with the boron-to-carbon measurements at the 95$\%$, 99$\%$ and 99.9$\%$ confidence levels, respectively. We also show in each frame the result using a more extreme parameter value, incompatible with the measured boron-to-carbon ratio. 

Previous authors have suggested that the observed titanium-to-iron ratio and/or the positron fraction could be explained for parameter values of $n^{-}_{gas}=2$ cm$^{-3}$ and $K_{B}=40$~\cite{Mertsch:2009ph} or 20~\cite{Blasi:2009hv} (and with the same values of $B$, $v^{-}$, and $r$ used here). It is clear from Fig.~\ref{fig:B_to_C}, however, that such models would also predict a very evident rise in the boron-to-carbon ratio, incompatible with the measured spectrum at well beyond the 99.9\% confidence level. 

As described in Sec.~\ref{sec:set_up}, we include the CR primary, CR accelerated secondary, and CR ISM secondary components for each CR species (and for their stable and long-lived isotopes). 
For the same parameters considered in Fig.~\ref{fig:B_to_C}, we predict that other secondary-to-primary ratios will also rise with energy, including the antiproton-to-proton and positron-to-electron ratios.
In Fig.~\ref{fig:pbar_to_p}, we show the CR antiproton-to-proton ratio predicted for some of the parameter values used in Fig.~\ref{fig:B_to_C}. There are significant uncertainties pertaining to the expected antiproton flux at high energies, $E > 50$ GeV, arising in part due to uncertainties in the cross section for antiproton production in p-p collisions (GALPROP uses cross sections as described in Ref.~\cite{Tan:1983de}, whereas we instead follow Ref.~\cite{Duperray:2003bd}). 
%
%
Given this uncertainty, a wide range of parameters for the acceleration of secondaries in SNRs could be potentially compatible with the observed antiproton-to-proton ratio. Yet, in all cases considered, parameter values of $K_{B}=40$, $n^{-}_{gas}$=2 cm$^{-3}$ are in considerable tension with the data. Reducing $K_{B}$ to 20, as in Ref.~\cite{Blasi:2009bd}, alleviates most of this tension, however. With future data extending to higher energies, such as that anticipated from \textit{AMS}, it will be possible to constrain such scenarios much more tightly. We also note as it has been shown in connection with dark matter annihilations/decays in the Galaxy \cite{Cirelli:2008pk, Donato:2008jk, Cholis:2010xb}, 
that the simple GALPROP (or semi-analytically calculated) background model for the antiproton-to-proton ratio agrees very well with the data, suggesting no evidence of an excess at high energies.

\begin{figure*}
\begin{centering}
\hspace{-0.5cm}
\includegraphics[width=3.40in,angle=0]{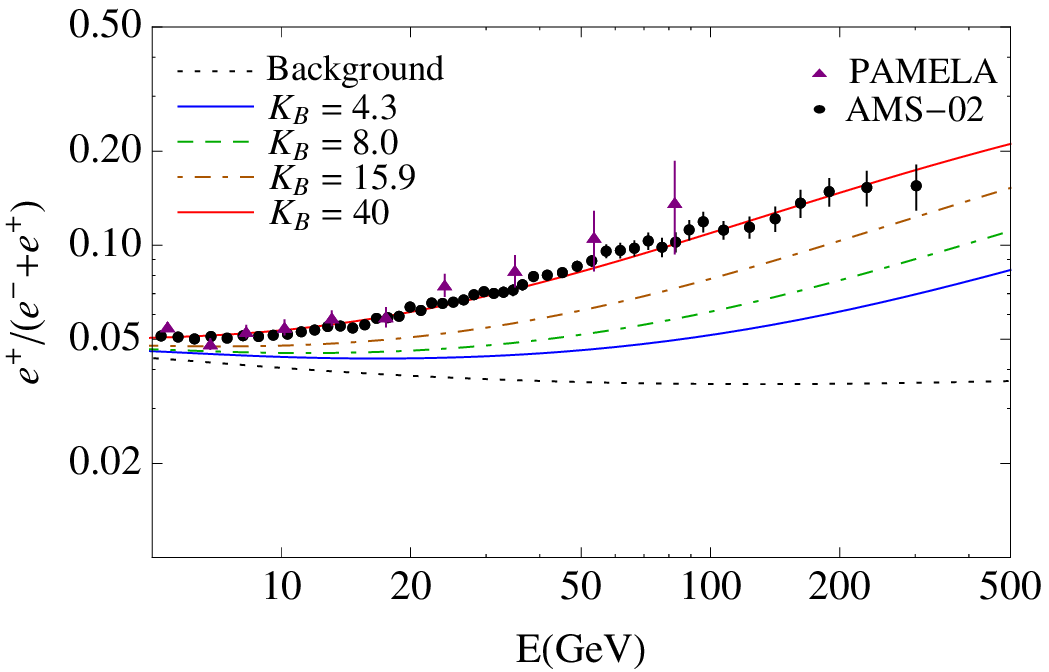}
\includegraphics[width=3.40in,angle=0]{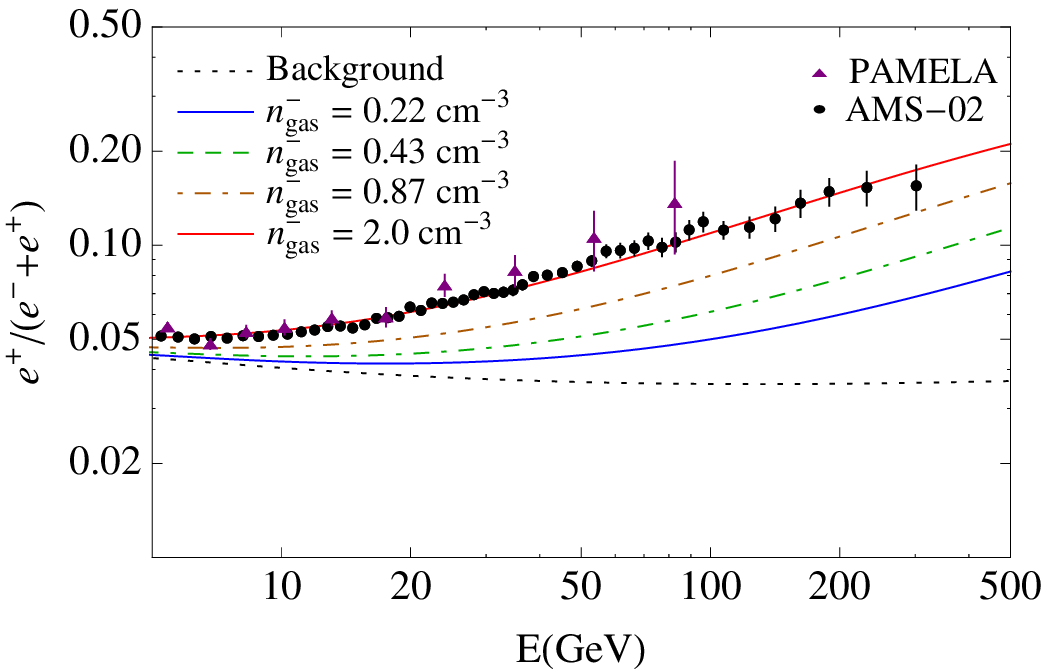} \end{centering}
\caption{The cosmic ray positron fraction predicted for the parameter choices used in Figs.~\ref{fig:B_to_C} and~\ref{fig:pbar_to_p}. The dotted black line denotes the prediction from secondary positrons produced in the interstellar medium (see Eqs.~\ref{eq:ISM_Spect_PrimElec} and~\ref{eq:ISM_Spect_SecPos}), without any contribution from positrons accelerated in the shocks of supernova remnants.  In the left frame, we set $n^{-}_{gas}$ = 2 cm$^{-3}$ and vary the value of $K_B$. In the right frame, we set $K_{B}=40$ and consider a range of values for $n^{-}_{gas}$. In both frames, we take $B$ = 1 $\mu$G, $v^{-}$ = 0.5$\times 10^{8}$ cm s$^{-1}$ and $r$ = 4 up to the highest energies. We also take $E_{max}$ = 10 TeV (see Fig.~\ref{fig:pos_frac2}). Although the measured positron fraction can be accommodated by a model with $K_B=40$ and $n^-_{gas}$=2 cm$^{-3}$, such a scenario is highly incompatible with the measured boron-to-carbon ratio  (and, to a lesser degree, with the antiproton-to-proton ratio). If we limit ourselves to parameter choices that are compatible with boron-to-carbon at the 95\% confidence level, we find that the acceleration of secondary positrons in SNRs can account for only $\sim$25\% of the excess positrons observed above 30 GeV.}
\label{fig:pos_frac}
\end{figure*}

In Fig.~\ref{fig:pos_frac}, we show the positron fraction predicted for the parameter values used in Figs.~\ref{fig:B_to_C} and ~\ref{fig:pbar_to_p}. While we find that the measured positron fraction can be accommodated in models with rather extreme parameter choices ($K_B=40$ and $n^-_{gas}$=2 cm$^{-3}$), those parameters also predict a boron-to-carbon ratio (and, to a lesser degree, a antiproton-to-proton ratio) that is highly incompatible with measurements (see Fig.~\ref{fig:B_to_C}). If we limit ourselves to parameter choices that are compatible with the boron-to-carbon ratio (at the 95\% confidence level, for example), we find that the acceleration of secondary positrons in SNRs can account for only $\sim$25\% of the excess positrons observed above 30 GeV. 

\begin{figure}
\begin{centering}
\includegraphics[width=3.40in,angle=0]{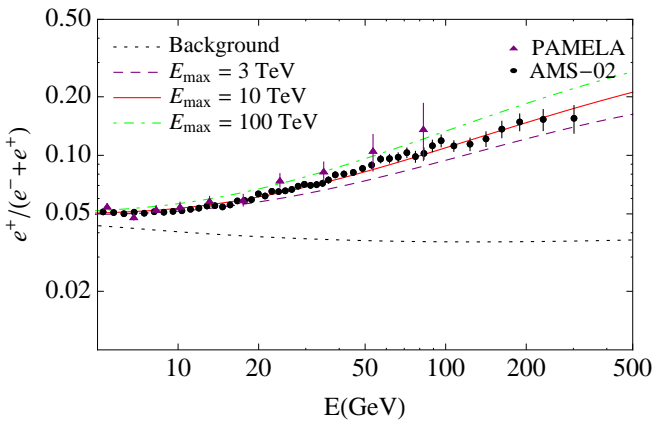}
\end{centering}
\caption{The impact of varying the value of the maximum energy to which $e^{\pm}$ are accelerated inside of supernova remnants. See text for details.}
\label{fig:pos_frac2}
\end{figure}

In producing Fig.~\ref{fig:pos_frac}, we have adopted a value of 10 TeV for $E_{max}$, the maximum energy to which secondary positrons are accelerated inside of SNRs (see Eq.~\ref{eq:ISM_Spect_SecPos}). In Fig.~\ref{fig:pos_frac2}, we show the impact of varying this quantity. We find that allowing for a higher value of $E_{max}\sim 100$ TeV can enable a value of $K_{B} \simeq 20$ to explain the rise of the positron fraction (with $n^{-}_{gas} = 2$ cm$^{-3}$). Even this somewhat lower value, however, predicts a boron-to-carbon ratio that is excluded at beyond the 99.9\% confidence level. 


In our calculations of the positron fraction, we have neglected energy losses from synchrotron emission and inverse Compton scattering inside of the SNRs (as had the authors of Refs.~\cite{Blasi:2009hv, Mertsch:2009ph}).\footnote{We have of course included energy losses during propagation in the ISM.} As a consequence, our results presented here are conservative, in that additional energy losses would only serve to further soften the contribution from accelerated positron secondaries. 
Furthermore, we note that we have allowed for the background positron fraction
to be practically flat in energy, giving the maximal contribution to the positron fraction at high energies.  

The limits presented here are quite robust. In particular, the compression ratio $r$ must be close to 4 in order for SNRs to be efficient accelerators, yielding a hard spectrum. Furthermore, the parameters  $K_{B}$, $B$, $n^{-}_{gas}$, $v^{-}$ are each interconnected, and the curves presented in Figs.~\ref{fig:B_to_C},~\ref{fig:pbar_to_p}, and~\ref{fig:pos_frac}, increase and decrease in the same manner with their variation as with $K_{B}$ and $n^{-}_{gas}$.  

\section{Discussion and Summary}
\label{sec:Conclusions}

The Galactic cosmic ray (CR) spectrum can be broken into two major components: primaries which are directly accelerated by supernova remnants, and secondaries which are produced in the interstellar medium from the spallation or decay of other CRs. In addition, secondaries that are produced inside of supernova remnants could be further accelerated before escaping.  Of particular interest is the possibility that these accelerated secondaries could account for the rising CR positron fraction, as measured by the \textit{PAMELA}, \textit{Fermi} and \textit{AMS} experiments~\cite{Blasi:2009hv, Mertsch:2009ph, Ahlers:2009ae}. In this paper, we revisit this scenario in light of recent CR data, such as \textit{AMS}'s measurement of the boron-to-carbon ratio. 

In agreement with previous groups~\cite{Blasi:2009hv, Mertsch:2009ph, Ahlers:2009ae}, we find that accelerated secondaries could plausibly lead to a positron fraction that rises with energy. Such models, however, also predict a significant rise in other secondary-to-primary ratios, which we find to be incompatible with recent observations. In particular, measurements of the boron-to-carbon and antiproton-to-proton ratios from the \textit{AMS} and \textit{PAMELA} collaborations show no evidence of a rise. 
We have tested different models by changing the gas density of the medium around the shock ($n^{-}_{gas}$) 
and the efficiency of diffusion of CRs up-stream and down-stream from the shock front ($K_B$). 
Quantitatively, we find that the observed boron-to-carbon ratio is incompatible (at the 95\% confidence level) with models in which more than $\sim$$25\%$ of the high energy excess positrons are secondaries that were produced and accelerated in the shocks of supernova remnants.


The constraints presented in this paper could be mitigated to some extent if different CR species were to originate from different sources.  For example, one could imagine a scenario in which the CR positrons were largely produced (as accelerated secondaries) in a few nearby supernova remnants (with high values of $K_B$ and/or $n^-_{gas}$). If the environments of those particular supernova remnants contained exceptionally high ambient densities of light nuclei (p, He), their relative contribution to the spectra of heavier CR nuclei producing boron could be suppressed. The boron, carbon and heavier nuclei composition of the CR spectrum would thus be set by other, more distant SNRs, perhaps less efficient in accelerating secondaries. While one can debate the plausibility of such a scenario, it is at least possible, in principle, to break the connection between the predicted positron fraction and boron-to-carbon ratio in this way. The connection between the positron fraction and the antiproton-to-proton ratio, however, cannot be broken in such a manner. If the \textit{AMS} experiment does not detect a significant rise in the high energy antiproton-to-proton ratio, that would likely rule out any remaining possibility that the rising positron fraction results from the acceleration of CR secondaries. 


\vskip 0.2 in
\section*{Acknowledgments}  
We would like to thank Mirko Boezio for valuable discussions. 
This work has been supported by the US Department of Energy. We would like to thank also the Aspen Center for Physics and the NSF Grant \#1066293 for hospitality during the earlier stages of this project.

\vskip 0.05in

\bibliography{BtoC_StochasticAccel}

\begin{thebibliography}{70}
\expandafter\ifx\csname natexlab\endcsname\relax\def\natexlab#1{#1}\fi
\expandafter\ifx\csname bibnamefont\endcsname\relax
  \def\bibnamefont#1{#1}\fi
\expandafter\ifx\csname bibfnamefont\endcsname\relax
  \def\bibfnamefont#1{#1}\fi
\expandafter\ifx\csname citenamefont\endcsname\relax
  \def\citenamefont#1{#1}\fi
\expandafter\ifx\csname url\endcsname\relax
  \def\url#1{\texttt{#1}}\fi
\expandafter\ifx\csname urlprefix\endcsname\relax\def\urlprefix{URL }\fi
\providecommand{\bibinfo}[2]{#2}
\providecommand{\eprint}[2][]{\url{#2}}

\bibitem[{\citenamefont{Aguilar et~al.}(2013)}]{AMS02}
\bibinfo{author}{\bibfnamefont{M.}~\bibnamefont{Aguilar}} \bibnamefont{et~al.}
  (\bibinfo{collaboration}{AMS Collaboration}), \bibinfo{journal}{Phys. Rev.
  Lett.} \textbf{\bibinfo{volume}{110}}, \bibinfo{pages}{141102}
  (\bibinfo{year}{2013}),
  \urlprefix\url{http://link.aps.org/doi/10.1103/PhysRevLett.110.141102}.

\bibitem[{\citenamefont{Adriani et~al.}(2009)}]{Adriani:2008zr}
\bibinfo{author}{\bibfnamefont{O.}~\bibnamefont{Adriani}} \bibnamefont{et~al.}
  (\bibinfo{collaboration}{PAMELA}), \bibinfo{journal}{Nature}
  \textbf{\bibinfo{volume}{458}}, \bibinfo{pages}{607} (\bibinfo{year}{2009}),
  \eprint{0810.4995}.

\bibitem[{\citenamefont{Picozza et~al.}(2007)}]{Picozza:2006nm}
\bibinfo{author}{\bibfnamefont{P.}~\bibnamefont{Picozza}} \bibnamefont{et~al.},
  \bibinfo{journal}{Astropart. Phys.} \textbf{\bibinfo{volume}{27}},
  \bibinfo{pages}{296} (\bibinfo{year}{2007}), \eprint{astro-ph/0608697}.

\bibitem[{\citenamefont{Gehrels and Michelson}(1999)}]{Gehrels:1999ri}
\bibinfo{author}{\bibfnamefont{N.}~\bibnamefont{Gehrels}} \bibnamefont{and}
  \bibinfo{author}{\bibfnamefont{P.}~\bibnamefont{Michelson}},
  \bibinfo{journal}{Astropart. Phys.} \textbf{\bibinfo{volume}{11}},
  \bibinfo{pages}{277} (\bibinfo{year}{1999}).

\bibitem[{\citenamefont{Bergstrom et~al.}(2008)\citenamefont{Bergstrom,
  Bringmann, and Edsjo}}]{Bergstrom:2008gr}
\bibinfo{author}{\bibfnamefont{L.}~\bibnamefont{Bergstrom}},
  \bibinfo{author}{\bibfnamefont{T.}~\bibnamefont{Bringmann}},
  \bibnamefont{and} \bibinfo{author}{\bibfnamefont{J.}~\bibnamefont{Edsjo}},
  \bibinfo{journal}{Phys. Rev.} \textbf{\bibinfo{volume}{D78}},
  \bibinfo{pages}{103520} (\bibinfo{year}{2008}), \eprint{0808.3725}.

\bibitem[{\citenamefont{Cirelli and Strumia}(2008)}]{Cirelli:2008jk}
\bibinfo{author}{\bibfnamefont{M.}~\bibnamefont{Cirelli}} \bibnamefont{and}
  \bibinfo{author}{\bibfnamefont{A.}~\bibnamefont{Strumia}},
  \bibinfo{journal}{PoS} \textbf{\bibinfo{volume}{IDM2008}},
  \bibinfo{pages}{089} (\bibinfo{year}{2008}), \eprint{0808.3867}.

\bibitem[{\citenamefont{Cholis et~al.}(2009{\natexlab{a}})\citenamefont{Cholis,
  Goodenough, Hooper, Simet, and Weiner}}]{Cholis:2008hb}
\bibinfo{author}{\bibfnamefont{I.}~\bibnamefont{Cholis}},
  \bibinfo{author}{\bibfnamefont{L.}~\bibnamefont{Goodenough}},
  \bibinfo{author}{\bibfnamefont{D.}~\bibnamefont{Hooper}},
  \bibinfo{author}{\bibfnamefont{M.}~\bibnamefont{Simet}}, \bibnamefont{and}
  \bibinfo{author}{\bibfnamefont{N.}~\bibnamefont{Weiner}},
  \bibinfo{journal}{Phys. Rev.} \textbf{\bibinfo{volume}{D80}},
  \bibinfo{pages}{123511} (\bibinfo{year}{2009}{\natexlab{a}}),
  \eprint{0809.1683}.

\bibitem[{\citenamefont{Barger et~al.}(2008)\citenamefont{Barger, Keung,
  Marfatia, and Shaughnessy}}]{Barger:2008su}
\bibinfo{author}{\bibfnamefont{V.}~\bibnamefont{Barger}},
  \bibinfo{author}{\bibfnamefont{W.~Y.} \bibnamefont{Keung}},
  \bibinfo{author}{\bibfnamefont{D.}~\bibnamefont{Marfatia}}, \bibnamefont{and}
  \bibinfo{author}{\bibfnamefont{G.}~\bibnamefont{Shaughnessy}}
  (\bibinfo{year}{2008}), \eprint{0809.0162}.

\bibitem[{\citenamefont{Cirelli et~al.}(2009)\citenamefont{Cirelli, Kadastik,
  Raidal, and Strumia}}]{Cirelli:2008pk}
\bibinfo{author}{\bibfnamefont{M.}~\bibnamefont{Cirelli}},
  \bibinfo{author}{\bibfnamefont{M.}~\bibnamefont{Kadastik}},
  \bibinfo{author}{\bibfnamefont{M.}~\bibnamefont{Raidal}}, \bibnamefont{and}
  \bibinfo{author}{\bibfnamefont{A.}~\bibnamefont{Strumia}},
  \bibinfo{journal}{Nucl.Phys.} \textbf{\bibinfo{volume}{B813}},
  \bibinfo{pages}{1} (\bibinfo{year}{2009}), \eprint{0809.2409}.

\bibitem[{\citenamefont{Nelson and Spitzer}(2010)}]{Nelson:2008hj}
\bibinfo{author}{\bibfnamefont{A.~E.} \bibnamefont{Nelson}} \bibnamefont{and}
  \bibinfo{author}{\bibfnamefont{C.}~\bibnamefont{Spitzer}},
  \bibinfo{journal}{JHEP} \textbf{\bibinfo{volume}{1010}}, \bibinfo{pages}{066}
  (\bibinfo{year}{2010}), \eprint{0810.5167}.

\bibitem[{\citenamefont{Arkani-Hamed et~al.}(2009)\citenamefont{Arkani-Hamed,
  Finkbeiner, Slatyer, and Weiner}}]{ArkaniHamed:2008qn}
\bibinfo{author}{\bibfnamefont{N.}~\bibnamefont{Arkani-Hamed}},
  \bibinfo{author}{\bibfnamefont{D.~P.} \bibnamefont{Finkbeiner}},
  \bibinfo{author}{\bibfnamefont{T.~R.} \bibnamefont{Slatyer}},
  \bibnamefont{and} \bibinfo{author}{\bibfnamefont{N.}~\bibnamefont{Weiner}},
  \bibinfo{journal}{Phys.Rev.} \textbf{\bibinfo{volume}{D79}},
  \bibinfo{pages}{015014} (\bibinfo{year}{2009}), \eprint{0810.0713}.

\bibitem[{\citenamefont{Cholis et~al.}(2009{\natexlab{b}})\citenamefont{Cholis,
  Finkbeiner, Goodenough, and Weiner}}]{Cholis:2008qq}
\bibinfo{author}{\bibfnamefont{I.}~\bibnamefont{Cholis}},
  \bibinfo{author}{\bibfnamefont{D.~P.} \bibnamefont{Finkbeiner}},
  \bibinfo{author}{\bibfnamefont{L.}~\bibnamefont{Goodenough}},
  \bibnamefont{and} \bibinfo{author}{\bibfnamefont{N.}~\bibnamefont{Weiner}},
  \bibinfo{journal}{JCAP} \textbf{\bibinfo{volume}{0912}}, \bibinfo{pages}{007}
  (\bibinfo{year}{2009}{\natexlab{b}}), \eprint{0810.5344}.

\bibitem[{\citenamefont{Nomura and Thaler}(2009)}]{Nomura:2008ru}
\bibinfo{author}{\bibfnamefont{Y.}~\bibnamefont{Nomura}} \bibnamefont{and}
  \bibinfo{author}{\bibfnamefont{J.}~\bibnamefont{Thaler}},
  \bibinfo{journal}{Phys.Rev.} \textbf{\bibinfo{volume}{D79}},
  \bibinfo{pages}{075008} (\bibinfo{year}{2009}), \eprint{0810.5397}.

\bibitem[{\citenamefont{Yin et~al.}(2009)\citenamefont{Yin, Yuan, Liu, Zhang,
  Bi et~al.}}]{Yin:2008bs}
\bibinfo{author}{\bibfnamefont{P.-f.} \bibnamefont{Yin}},
  \bibinfo{author}{\bibfnamefont{Q.}~\bibnamefont{Yuan}},
  \bibinfo{author}{\bibfnamefont{J.}~\bibnamefont{Liu}},
  \bibinfo{author}{\bibfnamefont{J.}~\bibnamefont{Zhang}},
  \bibinfo{author}{\bibfnamefont{X.-j.} \bibnamefont{Bi}},
  \bibnamefont{et~al.}, \bibinfo{journal}{Phys.Rev.}
  \textbf{\bibinfo{volume}{D79}}, \bibinfo{pages}{023512}
  (\bibinfo{year}{2009}), \eprint{0811.0176}.

\bibitem[{\citenamefont{Harnik and Kribs}(2009)}]{Harnik:2008uu}
\bibinfo{author}{\bibfnamefont{R.}~\bibnamefont{Harnik}} \bibnamefont{and}
  \bibinfo{author}{\bibfnamefont{G.~D.} \bibnamefont{Kribs}},
  \bibinfo{journal}{Phys.Rev.} \textbf{\bibinfo{volume}{D79}},
  \bibinfo{pages}{095007} (\bibinfo{year}{2009}), \eprint{0810.5557}.

\bibitem[{\citenamefont{Fox and Poppitz}(2009)}]{Fox:2008kb}
\bibinfo{author}{\bibfnamefont{P.~J.} \bibnamefont{Fox}} \bibnamefont{and}
  \bibinfo{author}{\bibfnamefont{E.}~\bibnamefont{Poppitz}},
  \bibinfo{journal}{Phys.Rev.} \textbf{\bibinfo{volume}{D79}},
  \bibinfo{pages}{083528} (\bibinfo{year}{2009}), \eprint{0811.0399}.

\bibitem[{\citenamefont{Pospelov and Ritz}(2009)}]{Pospelov:2008jd}
\bibinfo{author}{\bibfnamefont{M.}~\bibnamefont{Pospelov}} \bibnamefont{and}
  \bibinfo{author}{\bibfnamefont{A.}~\bibnamefont{Ritz}},
  \bibinfo{journal}{Phys.Lett.} \textbf{\bibinfo{volume}{B671}},
  \bibinfo{pages}{391} (\bibinfo{year}{2009}), \eprint{0810.1502}.

\bibitem[{\citenamefont{March-Russell and West}(2009)}]{MarchRussell:2008tu}
\bibinfo{author}{\bibfnamefont{J.~D.} \bibnamefont{March-Russell}}
  \bibnamefont{and} \bibinfo{author}{\bibfnamefont{S.~M.} \bibnamefont{West}},
  \bibinfo{journal}{Phys.Lett.} \textbf{\bibinfo{volume}{B676}},
  \bibinfo{pages}{133} (\bibinfo{year}{2009}), \eprint{0812.0559}.

\bibitem[{\citenamefont{Chang and Goodenough}(2011)}]{Chang:2011xn}
\bibinfo{author}{\bibfnamefont{S.}~\bibnamefont{Chang}} \bibnamefont{and}
  \bibinfo{author}{\bibfnamefont{L.}~\bibnamefont{Goodenough}},
  \bibinfo{journal}{Phys.Rev.} \textbf{\bibinfo{volume}{D84}},
  \bibinfo{pages}{023524} (\bibinfo{year}{2011}), \eprint{1105.3976}.

\bibitem[{\citenamefont{Hooper et~al.}(2008)\citenamefont{Hooper, Blasi, and
  Serpico}}]{Hooper:2008kg}
\bibinfo{author}{\bibfnamefont{D.}~\bibnamefont{Hooper}},
  \bibinfo{author}{\bibfnamefont{P.}~\bibnamefont{Blasi}}, \bibnamefont{and}
  \bibinfo{author}{\bibfnamefont{P.~D.} \bibnamefont{Serpico}}
  (\bibinfo{year}{2008}), \eprint{0810.1527}.

\bibitem[{\citenamefont{Yuksel et~al.}(2008)\citenamefont{Yuksel, Kistler, and
  Stanev}}]{Yuksel:2008rf}
\bibinfo{author}{\bibfnamefont{H.}~\bibnamefont{Yuksel}},
  \bibinfo{author}{\bibfnamefont{M.~D.} \bibnamefont{Kistler}},
  \bibnamefont{and} \bibinfo{author}{\bibfnamefont{T.}~\bibnamefont{Stanev}}
  (\bibinfo{year}{2008}), \eprint{0810.2784}.

\bibitem[{\citenamefont{Profumo}(2011)}]{Profumo:2008ms}
\bibinfo{author}{\bibfnamefont{S.}~\bibnamefont{Profumo}},
  \bibinfo{journal}{Central Eur.J.Phys.} \textbf{\bibinfo{volume}{10}},
  \bibinfo{pages}{1} (\bibinfo{year}{2011}), \eprint{0812.4457}.

\bibitem[{\citenamefont{Malyshev et~al.}(2009)\citenamefont{Malyshev, Cholis,
  and Gelfand}}]{Malyshev:2009tw}
\bibinfo{author}{\bibfnamefont{D.}~\bibnamefont{Malyshev}},
  \bibinfo{author}{\bibfnamefont{I.}~\bibnamefont{Cholis}}, \bibnamefont{and}
  \bibinfo{author}{\bibfnamefont{J.}~\bibnamefont{Gelfand}},
  \bibinfo{journal}{Phys. Rev.} \textbf{\bibinfo{volume}{D80}},
  \bibinfo{pages}{063005} (\bibinfo{year}{2009}), \eprint{0903.1310}.

\bibitem[{\citenamefont{Grasso et~al.}(2009)}]{Grasso:2009ma}
\bibinfo{author}{\bibfnamefont{D.}~\bibnamefont{Grasso}} \bibnamefont{et~al.}
  (\bibinfo{collaboration}{Fermi LAT Collaboration}),
  \bibinfo{journal}{Astropart.Phys.} \textbf{\bibinfo{volume}{32}},
  \bibinfo{pages}{140} (\bibinfo{year}{2009}), \eprint{0905.0636}.

\bibitem[{\citenamefont{Blasi}(2009)}]{Blasi:2009hv}
\bibinfo{author}{\bibfnamefont{P.}~\bibnamefont{Blasi}},
  \bibinfo{journal}{Phys.Rev.Lett.} \textbf{\bibinfo{volume}{103}},
  \bibinfo{pages}{051104} (\bibinfo{year}{2009}), \eprint{0903.2794}.

\bibitem[{\citenamefont{Mertsch and Sarkar}(2009)}]{Mertsch:2009ph}
\bibinfo{author}{\bibfnamefont{P.}~\bibnamefont{Mertsch}} \bibnamefont{and}
  \bibinfo{author}{\bibfnamefont{S.}~\bibnamefont{Sarkar}},
  \bibinfo{journal}{Phys.Rev.Lett.} \textbf{\bibinfo{volume}{103}},
  \bibinfo{pages}{081104} (\bibinfo{year}{2009}), \eprint{0905.3152}.

\bibitem[{\citenamefont{Ahlers et~al.}(2009)\citenamefont{Ahlers, Mertsch, and
  Sarkar}}]{Ahlers:2009ae}
\bibinfo{author}{\bibfnamefont{M.}~\bibnamefont{Ahlers}},
  \bibinfo{author}{\bibfnamefont{P.}~\bibnamefont{Mertsch}}, \bibnamefont{and}
  \bibinfo{author}{\bibfnamefont{S.}~\bibnamefont{Sarkar}},
  \bibinfo{journal}{Phys.Rev.} \textbf{\bibinfo{volume}{D80}},
  \bibinfo{pages}{123017} (\bibinfo{year}{2009}), \eprint{0909.4060}.

\bibitem[{\citenamefont{Ackermann et~al.}(2010)}]{Ackermann:2010ij}
\bibinfo{author}{\bibfnamefont{M.}~\bibnamefont{Ackermann}}
  \bibnamefont{et~al.} (\bibinfo{collaboration}{Fermi LAT Collaboration}),
  \bibinfo{journal}{Phys.Rev.} \textbf{\bibinfo{volume}{D82}},
  \bibinfo{pages}{092004} (\bibinfo{year}{2010}), \eprint{1008.3999}.

\bibitem[{\citenamefont{Abdo et~al.}(2009)}]{Abdo:2009zk}
\bibinfo{author}{\bibfnamefont{A.~A.} \bibnamefont{Abdo}} \bibnamefont{et~al.}
  (\bibinfo{collaboration}{Fermi LAT Collaboration}), \bibinfo{journal}{Phys.
  Rev. Lett.} \textbf{\bibinfo{volume}{102}}, \bibinfo{pages}{181101}
  (\bibinfo{year}{2009}), \eprint{0905.0025}.

\bibitem[{\citenamefont{Aharonian et~al.}(2008)}]{Collaboration:2008aaa}
\bibinfo{author}{\bibfnamefont{F.}~\bibnamefont{Aharonian}}
  \bibnamefont{et~al.} (\bibinfo{collaboration}{HESS Collaboration}),
  \bibinfo{journal}{Phys. Rev. Lett.} \textbf{\bibinfo{volume}{101}},
  \bibinfo{pages}{261104} (\bibinfo{year}{2008}), \eprint{0811.3894}.

\bibitem[{\citenamefont{Aharonian et~al.}(2009)}]{Aharonian:2009ah}
\bibinfo{author}{\bibfnamefont{F.}~\bibnamefont{Aharonian}}
  \bibnamefont{et~al.} (\bibinfo{collaboration}{HESS Collaboration}),
  \bibinfo{journal}{Astron. Astrophys.} \textbf{\bibinfo{volume}{508}},
  \bibinfo{pages}{561} (\bibinfo{year}{2009}), \eprint{0905.0105}.

\bibitem[{\citenamefont{Cholis and Hooper}(2013)}]{Cholis:2013psa}
\bibinfo{author}{\bibfnamefont{I.}~\bibnamefont{Cholis}} \bibnamefont{and}
  \bibinfo{author}{\bibfnamefont{D.}~\bibnamefont{Hooper}},
  \bibinfo{journal}{Phys.Rev.} \textbf{\bibinfo{volume}{D88}},
  \bibinfo{pages}{023013} (\bibinfo{year}{2013}), \eprint{1304.1840}.

\bibitem[{\citenamefont{Finkbeiner and Weiner}(2007)}]{Finkbeiner:2007kk}
\bibinfo{author}{\bibfnamefont{D.~P.} \bibnamefont{Finkbeiner}}
  \bibnamefont{and} \bibinfo{author}{\bibfnamefont{N.}~\bibnamefont{Weiner}},
  \bibinfo{journal}{Phys. Rev.} \textbf{\bibinfo{volume}{D76}},
  \bibinfo{pages}{083519} (\bibinfo{year}{2007}), \eprint{astro-ph/0702587}.

\bibitem[{\citenamefont{Cholis et~al.}(2009{\natexlab{c}})\citenamefont{Cholis,
  Dobler, Finkbeiner, Goodenough, and Weiner}}]{Cholis:2008wq}
\bibinfo{author}{\bibfnamefont{I.}~\bibnamefont{Cholis}},
  \bibinfo{author}{\bibfnamefont{G.}~\bibnamefont{Dobler}},
  \bibinfo{author}{\bibfnamefont{D.~P.} \bibnamefont{Finkbeiner}},
  \bibinfo{author}{\bibfnamefont{L.}~\bibnamefont{Goodenough}},
  \bibnamefont{and} \bibinfo{author}{\bibfnamefont{N.}~\bibnamefont{Weiner}},
  \bibinfo{journal}{Phys. Rev.} \textbf{\bibinfo{volume}{D80}},
  \bibinfo{pages}{123518} (\bibinfo{year}{2009}{\natexlab{c}}),
  \eprint{0811.3641}.

\bibitem[{\citenamefont{Chen et~al.}(2008)\citenamefont{Chen, Takahashi, and
  Yanagida}}]{Chen:2008qs}
\bibinfo{author}{\bibfnamefont{C.-R.} \bibnamefont{Chen}},
  \bibinfo{author}{\bibfnamefont{F.}~\bibnamefont{Takahashi}},
  \bibnamefont{and} \bibinfo{author}{\bibfnamefont{T.~T.}
  \bibnamefont{Yanagida}} (\bibinfo{year}{2008}), \eprint{0811.3357}.

\bibitem[{\citenamefont{Nardi et~al.}(2009)\citenamefont{Nardi, Sannino, and
  Strumia}}]{Nardi:2008ix}
\bibinfo{author}{\bibfnamefont{E.}~\bibnamefont{Nardi}},
  \bibinfo{author}{\bibfnamefont{F.}~\bibnamefont{Sannino}}, \bibnamefont{and}
  \bibinfo{author}{\bibfnamefont{A.}~\bibnamefont{Strumia}},
  \bibinfo{journal}{JCAP} \textbf{\bibinfo{volume}{0901}}, \bibinfo{pages}{043}
  (\bibinfo{year}{2009}), \eprint{0811.4153}.

\bibitem[{\citenamefont{Lattanzi and Silk}(2009)}]{Lattanzi:2008qa}
\bibinfo{author}{\bibfnamefont{M.}~\bibnamefont{Lattanzi}} \bibnamefont{and}
  \bibinfo{author}{\bibfnamefont{J.~I.} \bibnamefont{Silk}},
  \bibinfo{journal}{Phys.Rev.} \textbf{\bibinfo{volume}{D79}},
  \bibinfo{pages}{083523} (\bibinfo{year}{2009}), \eprint{0812.0360}.

\bibitem[{\citenamefont{Hisano et~al.}(2005)\citenamefont{Hisano, Matsumoto,
  Nojiri, and Saito}}]{Hisano:2004ds}
\bibinfo{author}{\bibfnamefont{J.}~\bibnamefont{Hisano}},
  \bibinfo{author}{\bibfnamefont{S.}~\bibnamefont{Matsumoto}},
  \bibinfo{author}{\bibfnamefont{M.~M.} \bibnamefont{Nojiri}},
  \bibnamefont{and} \bibinfo{author}{\bibfnamefont{O.}~\bibnamefont{Saito}},
  \bibinfo{journal}{Phys. Rev.} \textbf{\bibinfo{volume}{D71}},
  \bibinfo{pages}{063528} (\bibinfo{year}{2005}), \eprint{hep-ph/0412403}.

\bibitem[{\citenamefont{Kamionkowski et~al.}(2010)\citenamefont{Kamionkowski,
  Koushiappas, and Kuhlen}}]{Kamionkowski:2010mi}
\bibinfo{author}{\bibfnamefont{M.}~\bibnamefont{Kamionkowski}},
  \bibinfo{author}{\bibfnamefont{S.~M.} \bibnamefont{Koushiappas}},
  \bibnamefont{and} \bibinfo{author}{\bibfnamefont{M.}~\bibnamefont{Kuhlen}},
  \bibinfo{journal}{Phys.Rev.} \textbf{\bibinfo{volume}{D81}},
  \bibinfo{pages}{043532} (\bibinfo{year}{2010}), \eprint{1001.3144}.

\bibitem[{\citenamefont{Dienes et~al.}(2013)\citenamefont{Dienes, Kumar, and
  Thomas}}]{Dienes:2013lxa}
\bibinfo{author}{\bibfnamefont{K.~R.} \bibnamefont{Dienes}},
  \bibinfo{author}{\bibfnamefont{J.}~\bibnamefont{Kumar}}, \bibnamefont{and}
  \bibinfo{author}{\bibfnamefont{B.}~\bibnamefont{Thomas}}
  (\bibinfo{year}{2013}), \eprint{1306.2959}.

\bibitem[{\citenamefont{Bergstrom et~al.}(2013)\citenamefont{Bergstrom,
  Bringmann, Cholis, Hooper, and Weniger}}]{Bergstrom:2013jra}
\bibinfo{author}{\bibfnamefont{L.}~\bibnamefont{Bergstrom}},
  \bibinfo{author}{\bibfnamefont{T.}~\bibnamefont{Bringmann}},
  \bibinfo{author}{\bibfnamefont{I.}~\bibnamefont{Cholis}},
  \bibinfo{author}{\bibfnamefont{D.}~\bibnamefont{Hooper}}, \bibnamefont{and}
  \bibinfo{author}{\bibfnamefont{C.}~\bibnamefont{Weniger}},
  \bibinfo{journal}{Phys.Rev.Lett.} \textbf{\bibinfo{volume}{111}},
  \bibinfo{pages}{171101} (\bibinfo{year}{2013}), \eprint{1306.3983}.

\bibitem[{\citenamefont{Ibarra et~al.}(2013)\citenamefont{Ibarra,
  Lamperstorfer, and Silk}}]{Ibarra:2013zia}
\bibinfo{author}{\bibfnamefont{A.}~\bibnamefont{Ibarra}},
  \bibinfo{author}{\bibfnamefont{A.~S.} \bibnamefont{Lamperstorfer}},
  \bibnamefont{and} \bibinfo{author}{\bibfnamefont{J.}~\bibnamefont{Silk}}
  (\bibinfo{year}{2013}), \eprint{1309.2570}.

\bibitem[{\citenamefont{Linden and Profumo}(2013)}]{Linden:2013mqa}
\bibinfo{author}{\bibfnamefont{T.}~\bibnamefont{Linden}} \bibnamefont{and}
  \bibinfo{author}{\bibfnamefont{S.}~\bibnamefont{Profumo}},
  \bibinfo{journal}{Astrophys.J.} \textbf{\bibinfo{volume}{772}},
  \bibinfo{pages}{18} (\bibinfo{year}{2013}), \eprint{1304.1791}.

\bibitem[{\citenamefont{Gaggero et~al.}(2013)\citenamefont{Gaggero, Maccione,
  Di~Bernardo, Evoli, and Grasso}}]{Gaggero:2013rya}
\bibinfo{author}{\bibfnamefont{D.}~\bibnamefont{Gaggero}},
  \bibinfo{author}{\bibfnamefont{L.}~\bibnamefont{Maccione}},
  \bibinfo{author}{\bibfnamefont{G.}~\bibnamefont{Di~Bernardo}},
  \bibinfo{author}{\bibfnamefont{C.}~\bibnamefont{Evoli}}, \bibnamefont{and}
  \bibinfo{author}{\bibfnamefont{D.}~\bibnamefont{Grasso}},
  \bibinfo{journal}{Phys. Rev. Lett. 111,} \textbf{\bibinfo{volume}{021102}}
  (\bibinfo{year}{2013}), \eprint{1304.6718}.

\bibitem[{\citenamefont{Blasi and Serpico}(2009)}]{Blasi:2009bd}
\bibinfo{author}{\bibfnamefont{P.}~\bibnamefont{Blasi}} \bibnamefont{and}
  \bibinfo{author}{\bibfnamefont{P.~D.} \bibnamefont{Serpico}},
  \bibinfo{journal}{Phys.Rev.Lett.} \textbf{\bibinfo{volume}{103}},
  \bibinfo{pages}{081103} (\bibinfo{year}{2009}), \eprint{0904.0871}.

\bibitem[{\citenamefont{Kachelriess et~al.}(2011)\citenamefont{Kachelriess,
  Ostapchenko, and Tomas}}]{Kachelriess:2011qv}
\bibinfo{author}{\bibfnamefont{M.}~\bibnamefont{Kachelriess}},
  \bibinfo{author}{\bibfnamefont{S.}~\bibnamefont{Ostapchenko}},
  \bibnamefont{and} \bibinfo{author}{\bibfnamefont{R.}~\bibnamefont{Tomas}},
  \bibinfo{journal}{Astrophys.J.} \textbf{\bibinfo{volume}{733}},
  \bibinfo{pages}{119} (\bibinfo{year}{2011}), \eprint{1103.5765}.

\bibitem[{\citenamefont{Kachelrieß and
  Ostapchenko}(2013)}]{Kachelriess:2012ag}
\bibinfo{author}{\bibfnamefont{M.}~\bibnamefont{Kachelrieß}} \bibnamefont{and}
  \bibinfo{author}{\bibfnamefont{S.}~\bibnamefont{Ostapchenko}},
  \bibinfo{journal}{Phys.Rev.} \textbf{\bibinfo{volume}{D87}},
  \bibinfo{pages}{047301} (\bibinfo{year}{2013}), \eprint{1211.1033}.

\bibitem[{\citenamefont{Carbone}(ICRC 2013, Rio de Janeiro)}]{PAMELABtoC}
\bibinfo{author}{\bibfnamefont{R.}~\bibnamefont{Carbone}},
  \emph{\bibinfo{title}{Galactic boron and carbon fluxes by the pamela
  experiment}} (\bibinfo{year}{ICRC 2013, Rio de Janeiro}).

\bibitem[{\citenamefont{AMS-02}(2013)}]{AMSsite}
\bibinfo{author}{\bibnamefont{AMS-02}},
  \bibinfo{journal}{http://www.ams02.org/,http://www.ams02.org
  /2013/07/new-ams-data6-bc-ratio-5-gv-to-500-gv/}  (\bibinfo{year}{2013}).

\bibitem[{\citenamefont{Engelmann et~al.}(1990)\citenamefont{Engelmann,
  Ferrando, Soutoul, Goret, and Juliusson}}]{Engelmann:1990zz}
\bibinfo{author}{\bibfnamefont{J.}~\bibnamefont{Engelmann}},
  \bibinfo{author}{\bibfnamefont{P.}~\bibnamefont{Ferrando}},
  \bibinfo{author}{\bibfnamefont{A.}~\bibnamefont{Soutoul}},
  \bibinfo{author}{\bibfnamefont{P.}~\bibnamefont{Goret}}, \bibnamefont{and}
  \bibinfo{author}{\bibfnamefont{E.}~\bibnamefont{Juliusson}},
  \bibinfo{journal}{Astron.Astrophys.} \textbf{\bibinfo{volume}{233}},
  \bibinfo{pages}{96} (\bibinfo{year}{1990}).

\bibitem[{\citenamefont{Silberberg and Tsao}(1990)}]{Silberberg:1990nj}
\bibinfo{author}{\bibfnamefont{R.}~\bibnamefont{Silberberg}} \bibnamefont{and}
  \bibinfo{author}{\bibfnamefont{C.}~\bibnamefont{Tsao}},
  \bibinfo{journal}{Phys.Rept.} \textbf{\bibinfo{volume}{191}},
  \bibinfo{pages}{351} (\bibinfo{year}{1990}).

\bibitem[{\citenamefont{Silberberg and Tsao}(1973)}]{Silberberg:1973jxa}
\bibinfo{author}{\bibfnamefont{R.}~\bibnamefont{Silberberg}} \bibnamefont{and}
  \bibinfo{author}{\bibfnamefont{C.~H.} \bibnamefont{Tsao}},
  \bibinfo{journal}{Astrophys.J.Suppl.} \textbf{\bibinfo{volume}{25}},
  \bibinfo{pages}{315} (\bibinfo{year}{1973}).

\bibitem[{\citenamefont{Lorimer}(2004)}]{LorimerV1}
\bibinfo{author}{\bibfnamefont{M.~S.} \bibnamefont{Lorimer}},
  \emph{\bibinfo{title}{High energy astrophysics, pp. 339, vol. 1}}
  (\bibinfo{year}{2004}).

\bibitem[{\citenamefont{Strong et~al.}(2011)\citenamefont{Strong, Moskalenko,
  Porter, Orlando, Diger, and Vladimirov}}]{Galprop1}
\bibinfo{author}{\bibfnamefont{A.~W.} \bibnamefont{Strong}},
  \bibinfo{author}{\bibfnamefont{I.~V.} \bibnamefont{Moskalenko}},
  \bibinfo{author}{\bibfnamefont{T.~A.} \bibnamefont{Porter}},
  \bibinfo{author}{\bibfnamefont{E.}~\bibnamefont{Orlando}},
  \bibinfo{author}{\bibfnamefont{S.~W.} \bibnamefont{Diger}}, \bibnamefont{and}
  \bibinfo{author}{\bibfnamefont{A.~E.} \bibnamefont{Vladimirov}},
  \emph{\bibinfo{title}{GALPROP Version 54: Explanatory Supplement}}
  (\bibinfo{year}{2011}).

\bibitem[{\citenamefont{http://galprop.stanford.edu/}()}]{website}
\bibinfo{author}{\bibnamefont{http://galprop.stanford.edu/}}.

\bibitem[{\citenamefont{http://dragon.hepforge.org}()}]{DRAGONweb}
\bibinfo{author}{\bibnamefont{http://dragon.hepforge.org}}.

\bibitem[{\citenamefont{Simet and Hooper}(2009)}]{Simet:2009ne}
\bibinfo{author}{\bibfnamefont{M.}~\bibnamefont{Simet}} \bibnamefont{and}
  \bibinfo{author}{\bibfnamefont{D.}~\bibnamefont{Hooper}},
  \bibinfo{journal}{JCAP} \textbf{\bibinfo{volume}{0908}}, \bibinfo{pages}{003}
  (\bibinfo{year}{2009}), \eprint{0904.2398}.

\bibitem[{\citenamefont{Trotta et~al.}(2011)\citenamefont{Trotta, Johannesson,
  Moskalenko, Porter, de~Austri et~al.}}]{Trotta:2010mx}
\bibinfo{author}{\bibfnamefont{R.}~\bibnamefont{Trotta}},
  \bibinfo{author}{\bibfnamefont{G.}~\bibnamefont{Johannesson}},
  \bibinfo{author}{\bibfnamefont{I.}~\bibnamefont{Moskalenko}},
  \bibinfo{author}{\bibfnamefont{T.}~\bibnamefont{Porter}},
  \bibinfo{author}{\bibfnamefont{R.~R.} \bibnamefont{de~Austri}},
  \bibnamefont{et~al.}, \bibinfo{journal}{Astrophys.J.}
  \textbf{\bibinfo{volume}{729}}, \bibinfo{pages}{106} (\bibinfo{year}{2011}),
  \eprint{1011.0037}.

\bibitem[{\citenamefont{Gleeson and Axford}(1968)}]{Gleeson:1968zza}
\bibinfo{author}{\bibfnamefont{L.}~\bibnamefont{Gleeson}} \bibnamefont{and}
  \bibinfo{author}{\bibfnamefont{W.}~\bibnamefont{Axford}},
  \bibinfo{journal}{Astrophys.J.} \textbf{\bibinfo{volume}{154}},
  \bibinfo{pages}{1011} (\bibinfo{year}{1968}).

\bibitem[{\citenamefont{{Strauss} et~al.}(2011)\citenamefont{{Strauss},
  {Potgieter}, {B{\"u}sching}, and {Kopp}}}]{2011ApJ...735...83S}
\bibinfo{author}{\bibfnamefont{R.~D.} \bibnamefont{{Strauss}}},
  \bibinfo{author}{\bibfnamefont{M.~S.} \bibnamefont{{Potgieter}}},
  \bibinfo{author}{\bibfnamefont{I.}~\bibnamefont{{B{\"u}sching}}},
  \bibnamefont{and} \bibinfo{author}{\bibfnamefont{A.}~\bibnamefont{{Kopp}}},
  \bibinfo{journal}{\apj} \textbf{\bibinfo{volume}{735}}, \bibinfo{eid}{83}
  (\bibinfo{year}{2011}).

\bibitem[{\citenamefont{{Strauss} et~al.}(2012)\citenamefont{{Strauss},
  {Potgieter}, {B{\"u}sching}, and {Kopp}}}]{2012Ap&SS.339..223S}
\bibinfo{author}{\bibfnamefont{R.~D.} \bibnamefont{{Strauss}}},
  \bibinfo{author}{\bibfnamefont{M.~S.} \bibnamefont{{Potgieter}}},
  \bibinfo{author}{\bibfnamefont{I.}~\bibnamefont{{B{\"u}sching}}},
  \bibnamefont{and} \bibinfo{author}{\bibfnamefont{A.}~\bibnamefont{{Kopp}}},
  \bibinfo{journal}{\apss} \textbf{\bibinfo{volume}{339}}, \bibinfo{pages}{223}
  (\bibinfo{year}{2012}).

\bibitem[{\citenamefont{Maccione}(2013)}]{Maccione:2012cu}
\bibinfo{author}{\bibfnamefont{L.}~\bibnamefont{Maccione}},
  \bibinfo{journal}{Phys.Rev.Lett.} \textbf{\bibinfo{volume}{110}},
  \bibinfo{pages}{081101} (\bibinfo{year}{2013}), \eprint{1211.6905}.

\bibitem[{\citenamefont{Ahn et~al.}(2008)}]{Ahn:2008my}
\bibinfo{author}{\bibfnamefont{H.~S.} \bibnamefont{Ahn}} \bibnamefont{et~al.},
  \bibinfo{journal}{Astropart. Phys.} \textbf{\bibinfo{volume}{30}},
  \bibinfo{pages}{133} (\bibinfo{year}{2008}), \eprint{0808.1718}.

\bibitem[{\citenamefont{{Swordy} et~al.}(1990)\citenamefont{{Swordy},
  {Mueller}, {Meyer}, {L'Heureux}, and {Grunsfeld}}}]{1990ApJ...349..625S}
\bibinfo{author}{\bibfnamefont{S.~P.} \bibnamefont{{Swordy}}},
  \bibinfo{author}{\bibfnamefont{D.}~\bibnamefont{{Mueller}}},
  \bibinfo{author}{\bibfnamefont{P.}~\bibnamefont{{Meyer}}},
  \bibinfo{author}{\bibfnamefont{J.}~\bibnamefont{{L'Heureux}}},
  \bibnamefont{and} \bibinfo{author}{\bibfnamefont{J.~M.}
  \bibnamefont{{Grunsfeld}}}, \bibinfo{journal}{\apj}
  \textbf{\bibinfo{volume}{349}}, \bibinfo{pages}{625} (\bibinfo{year}{1990}).

\bibitem[{\citenamefont{de~Nolfo et~al.}(2006)\citenamefont{de~Nolfo,
  Moskalenko, Binns, Christian, Cummings et~al.}}]{deNolfo:2006qj}
\bibinfo{author}{\bibfnamefont{G.~A.} \bibnamefont{de~Nolfo}},
  \bibinfo{author}{\bibfnamefont{I.}~\bibnamefont{Moskalenko}},
  \bibinfo{author}{\bibfnamefont{W.}~\bibnamefont{Binns}},
  \bibinfo{author}{\bibfnamefont{E.}~\bibnamefont{Christian}},
  \bibinfo{author}{\bibfnamefont{A.}~\bibnamefont{Cummings}},
  \bibnamefont{et~al.}, \bibinfo{journal}{Adv.Space Res.}
  \textbf{\bibinfo{volume}{38}}, \bibinfo{pages}{1558} (\bibinfo{year}{2006}),
  \eprint{astro-ph/0611301}.

\bibitem[{\citenamefont{Obermeier et~al.}(2011)\citenamefont{Obermeier, Ave,
  Boyle, Hoppner, Horandel et~al.}}]{Obermeier:2011wm}
\bibinfo{author}{\bibfnamefont{A.}~\bibnamefont{Obermeier}},
  \bibinfo{author}{\bibfnamefont{M.}~\bibnamefont{Ave}},
  \bibinfo{author}{\bibfnamefont{P.}~\bibnamefont{Boyle}},
  \bibinfo{author}{\bibfnamefont{C.}~\bibnamefont{Hoppner}},
  \bibinfo{author}{\bibfnamefont{J.}~\bibnamefont{Horandel}},
  \bibnamefont{et~al.}, \bibinfo{journal}{Astrophys.J.}
  \textbf{\bibinfo{volume}{742}}, \bibinfo{pages}{14} (\bibinfo{year}{2011}),
  \eprint{1108.4838}.

\bibitem[{\citenamefont{Tan and Ng}(1983)}]{Tan:1983de}
\bibinfo{author}{\bibfnamefont{L.}~\bibnamefont{Tan}} \bibnamefont{and}
  \bibinfo{author}{\bibfnamefont{L.}~\bibnamefont{Ng}},
  \bibinfo{journal}{J.Phys.} \textbf{\bibinfo{volume}{G9}},
  \bibinfo{pages}{227} (\bibinfo{year}{1983}).

\bibitem[{\citenamefont{Duperray et~al.}(2003)\citenamefont{Duperray, Huang,
  Protasov, and Buenerd}}]{Duperray:2003bd}
\bibinfo{author}{\bibfnamefont{R.}~\bibnamefont{Duperray}},
  \bibinfo{author}{\bibfnamefont{C.-Y.} \bibnamefont{Huang}},
  \bibinfo{author}{\bibfnamefont{K.}~\bibnamefont{Protasov}}, \bibnamefont{and}
  \bibinfo{author}{\bibfnamefont{M.}~\bibnamefont{Buenerd}},
  \bibinfo{journal}{Phys.Rev.} \textbf{\bibinfo{volume}{D68}},
  \bibinfo{pages}{094017} (\bibinfo{year}{2003}), \eprint{astro-ph/0305274}.

\bibitem[{\citenamefont{Donato et~al.}(2009)\citenamefont{Donato, Maurin, Brun,
  Delahaye, and Salati}}]{Donato:2008jk}
\bibinfo{author}{\bibfnamefont{F.}~\bibnamefont{Donato}},
  \bibinfo{author}{\bibfnamefont{D.}~\bibnamefont{Maurin}},
  \bibinfo{author}{\bibfnamefont{P.}~\bibnamefont{Brun}},
  \bibinfo{author}{\bibfnamefont{T.}~\bibnamefont{Delahaye}}, \bibnamefont{and}
  \bibinfo{author}{\bibfnamefont{P.}~\bibnamefont{Salati}},
  \bibinfo{journal}{Phys. Rev. Lett.} \textbf{\bibinfo{volume}{102}},
  \bibinfo{pages}{071301} (\bibinfo{year}{2009}), \eprint{0810.5292}.

\bibitem[{\citenamefont{Cholis}(2011)}]{Cholis:2010xb}
\bibinfo{author}{\bibfnamefont{I.}~\bibnamefont{Cholis}},
  \bibinfo{journal}{JCAP} \textbf{\bibinfo{volume}{1109}}, \bibinfo{pages}{007}
  (\bibinfo{year}{2011}), \eprint{1007.1160}.

\end{thebibliography}
\bibliographystyle{apsrev}

\end{document}